\documentclass[aps,pra,superscriptaddress,floatfix,nofootinbib]{revtex4}
\usepackage{mathrsfs}
\usepackage{epsfig,psfrag}
\usepackage{amsmath,amsfonts,amssymb, latexsym}
\usepackage[usenames]{color}
\usepackage{subfigure}
\usepackage[vflt]{floatflt}
\usepackage{bm}
\usepackage{sidecap}

\def\tr{{\rm tr\,}}

\newcommand{\curl}{\boldsymbol{\nabla}\times}
\newcommand{\curlp}{\boldsymbol{\nabla}'\times}

\newcommand{\vecr}{\mathbf{r}}

\newcommand{\x}{\mathbf{r}}

\definecolor{BrickRed}{cmyk}{0,0.89,0.94,0.28}
\definecolor{MidnightBlue}{cmyk}{0.98,0.13,0,0.43}
\definecolor{DarkGreen}{rgb}{0,0.7,0.1}
\definecolor{te}{rgb}{0.92,0.,0.65}
\definecolor{sjr}{rgb}{0.5,0.3,1.0}
\definecolor{Green}{rgb}{0,1,0}
\definecolor{mfm}{rgb}{.8,.08,.05}

\begin{document}
\title{Casimir force between sharp--shaped conductors}
\author{Mohammad F. Maghrebi}
\affiliation{Center for Theoretical Physics, Laboratory for Nuclear Science, and Department of Physics, Massachusetts Institute of Technology, Cambridge, MA 02139, USA}
\affiliation{Department of Physics, Massachusetts Institute of Technology, Cambridge, MA 02139, USA}
\author{Sahand Jamal Rahi\footnote{Present address: Center for Studies in Physics and Biology, The Rockefeller University,
1230 York Street, New York, NY 10065, USA.}}
\affiliation{Department of Physics, Massachusetts Institute of Technology, Cambridge, MA 02139, USA}
\author{Thorsten Emig}
\affiliation{Laboratoire de Physique Th\'eorique et Mod\`eles Statistiques,  Universit\'e Paris-Sud, 91405 Orsay, France}
\author{Noah Graham}
\affiliation{Department of Physics, Middlebury College, Middlebury, VT 05753, USA}
\author{Robert L. Jaffe}
\affiliation{Center for Theoretical Physics, Laboratory for Nuclear Science, and Department of Physics, Massachusetts Institute of Technology, Cambridge, MA 02139, USA}
\author{Mehran Kardar}
\affiliation{Department of Physics, Massachusetts Institute of Technology, Cambridge, MA 02139, USA}

\begin{abstract}
Casimir forces between conductors at the sub-micron
scale cannot be ignored in the design and operation of micro-electromechanical (MEM) devices. However, these forces depend non-trivially on geometry, and existing formulae and
approximations cannot deal with realistic micro-machinery components with sharp edges and tips.
Here, we employ a novel approach to electromagnetic scattering,
appropriate to perfect conductors with sharp edges and tips,
specifically to wedges and cones. The interaction of these objects
with a metal plate (and among themselves) is then computed
systematically by a multiple-scattering series.
For the wedge, we obtain {\em analytical} expressions for the
interaction with a plate, as functions of opening angle and tilt,
which should provide a particularly useful tool for the design of
MEMs.
Our result for the Casimir interactions between conducting cones and
plates applies directly to the force on the tip of a scanning
tunneling probe; the unexpectedly large temperature dependence of the
force in these configurations should attract immediate experimental
interest.
\end{abstract}
\date \today
\maketitle

\section{Introduction}

The inherent appeal of the Casimir force as a macroscopic manifestation of quantum ``zero-point'' fluctuations has inspired many studies over the decades that followed its discovery \cite{Milton01}.  Casimir's original result \cite{Casimir48-2} for the force between perfectly reflecting mirrors separated by vacuum was quickly extended to include slabs of material with specified (frequency-dependent) dielectric response \cite{Dzyaloshinskii61}. Precise experimental confirmation, however, had to await the advent of high precision scanning probes \cite{Lamoreaux97,Mohideen98,Bressi02,Bordag09}.  Recent studies have aimed to reduce or reverse the attractive Casimir force in practical applications.  In the presence of an intervening fluid, experiments have indeed observed repulsion due to quantum \cite{Munday09} or critical thermal \cite{Hertlein08} fluctuations.  Metamaterials, fabricated designs of microcircuitry, have also been proposed as candidates for Casimir repulsion across an intervening vacuum \cite{Leonhardt07}.

Compared to studies of materials, the treatment of shapes and geometry has remained at a primitive stage.  Interactions between non-planar shapes are typically calculated via the {\it proximity force approximation} (PFA), which sums over infinitesimal segments treated as locally parallel plates \cite{Derjaguin56}. This is a serious limitation since the majority of experiments measure the force between a sphere and a plate, with precision that is now sufficient to probe deviations from PFA in this and other geometries \cite{Chan08,Decca07-1}.  Practical applications are likely to explore geometries further removed from parallel plates.

The formalism recently implemented in Refs.~\cite{Rahi09,Emig07} enables systematic computations of electromagnetic Casimir forces in terms of a multipole expansion.  Using these methods we have been able to compute electromagnetic Casimir forces between various combinations of planes, spheres, and circular and parabolic cylinders \cite{Emig07, Rahi08-1,Rahi08-2,Emig08,Graham10, Zandi10} (see also \cite{Kenneth08, Maia_Neto08, Canaguier-Durand10}). Both perfect conductors and dielectrics have been studied. However, with the notable exception of the knife-edge \cite{Graham10}, which is a limit of the parabolic cylinder geometry, systems with sharp edges have not yet been studied\footnote{Knife edge geometries have been studied for scalar fields obeying Dirichlet boundary conditions in Refs.~\cite{Gies06-2,Kabat10}. Wedges and related shapes have been studied in {\it isolation} \cite{Ellingsen10}, but computing interactions {\it amongst} such objects requires full scattering and conversion matrices, which are first synthesized in this paper.}.

In all the above cases, the object corresponds to a surface of constant {\it radial} coordinate.  {In this paper we present the first results on the quantum and thermal electromagnetic Casimir forces between generically sharp shapes, such as a wedge and a cone, and a conducting plane.}  {We accomplish this by considering surfaces of constant {\it angular} coordinate. While}  the conceptual step --- radial to angular --- is simple, the practical computation of scattering properties is nontrivial, necessitating complex, and in places novel, mathematical steps.  Furthermore, the inclusion of wedges and cones practically exhausts shapes for which the EM scattering amplitude {can be treated analytically}\footnote{Morse and Feshbach \cite{Morse53} enumerate six coordinate systems in which the vector Helmholtz equation for EM waves is generically solvable in this way: planar, cylindrical (comprising circular, elliptic, and parabolic), spherical, and conical.  Surfaces on which one such coordinate is constant are candidates for exact analysis.  In addition to the plane and sphere discussed above, the circular \cite{Emig06} (bottom-right of Fig.~\ref{elements}) and parabolic \cite{Graham10} cylinder have already been studied.  These shapes are generically smooth, with the extreme limit of the parabolic cylinder (a ``knife-edge'') a notable exception.  A survey of the remaining coordinate systems \cite{Morse53} only leads to shapes such as cylinders and cones with elliptic cross-section, which are generically similar to their circular counterparts.}.

{In light of the technical complexity of the analysis, we first describe our methods qualitatively and summarize our results in the next section.  The rest of the paper is organized as follows:  In Section~\ref{sec:wedge} we present the analysis for wedge geometries, where for perfect conductors the electromagnetic field can be parameterized in terms of two scalar fields.  In Section~\ref{sec:cone} we treat the cone, both for scalar fields and for electromagnetism, where the vector nature of the field is unavoidable.  In each case we not only describe the formalism but also present further figures and results.  The role of thermal fluctuations is discussed in Section~\ref{sec:thermal}, where we provide the explicit formulae for Casimir forces at finite
temperature. In Sec.~\ref{sec:conductivity}  we argue that the imperfect conductivity of
typical metals leads to controllably small corrections to the computed
Casimir forces. Finally, in an Appendix we discuss the (novel) representation of
the electromagnetic Green's function that we employ for computations of
scattering from a cone.}

\section{Overview and discussion}
\label{sec:results}
\begin{figure}[b]
\begin{center}
\includegraphics[width=13cm]{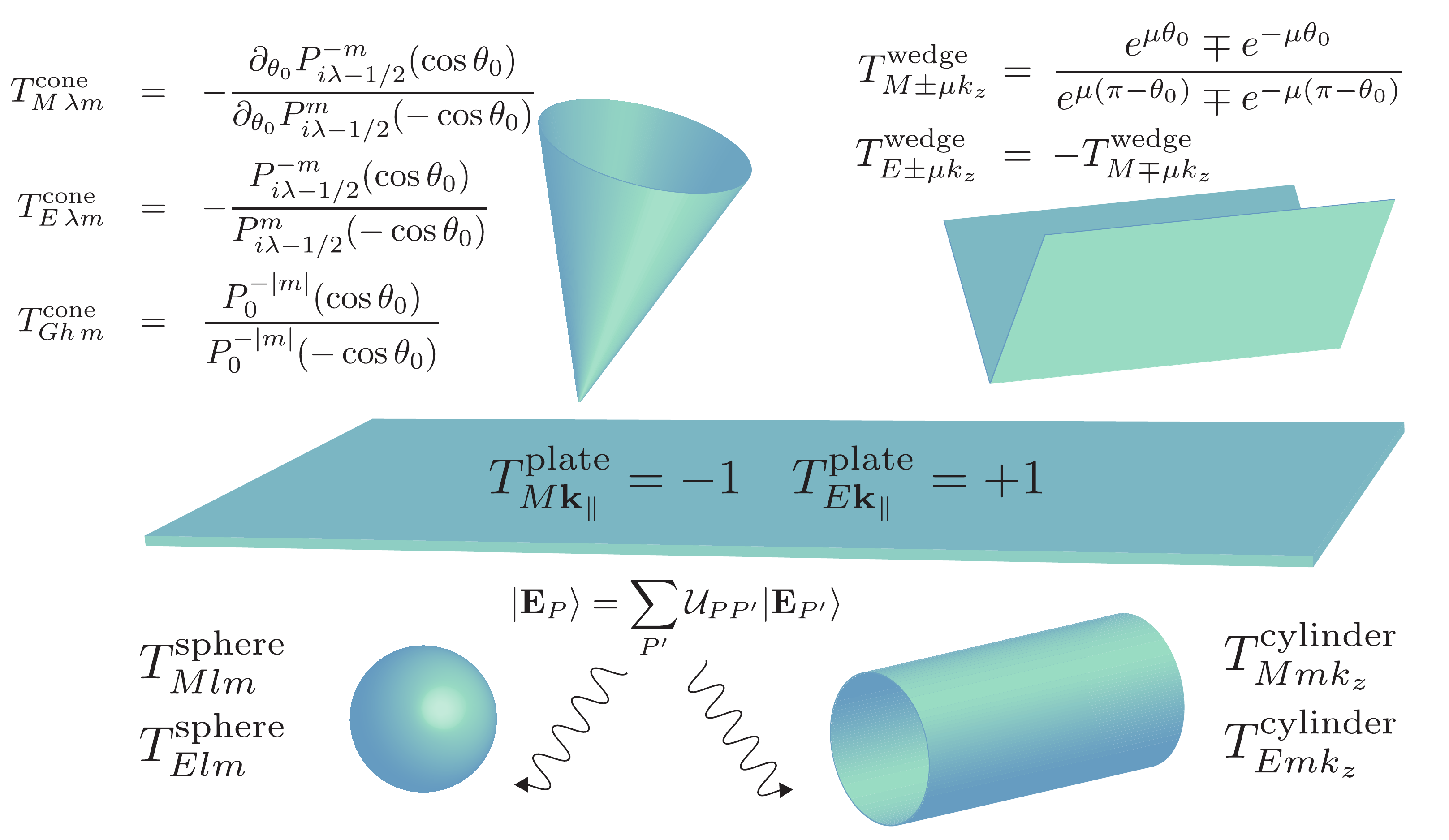}
\caption{{Ingredients in the scattering theory approach to electromagnetic Casimir forces.  See the text for further discussion.}}
\label{elements}
\end{center}
\end{figure}

The conceptual foundations of {the scattering} approach can be traced back to earlier multiple-scattering formalisms \cite{Balian77,Balian78,Kenneth06,Bordag09}, but these were not sufficiently transparent to enable practical calculations.  The ingredients in our method are depicted in Fig.~\ref{elements}. {The Casimir energy associated with a specific geometry depends on the way that the objects  {\it constrain} the electromagnetic waves that can bounce back and forth between them.   The dependence on the properties of the objects is completely encoded in the scattering amplitude or $T$-matrix for EM waves.  The $T$-matrices are indexed in a coordinate-basis suitable to each object, and by polarization.}  Fig.~1 summarizes the $T$-matrices for perfectly reflecting planar, wedge, and conical geometries.  Scattering from a plane mirror gives $T^{\rm plate}=\pm1$ {for the two polarizations}, irrespective of the wavevector $\vec k$. Cylindrical $(m,k_z)$ and spherical ($\ell,m$) quantum numbers label appropriate bases for the cylinder and sphere.
As detailed in Sections~\ref{sec:wedge} and \ref{sec:cone}, imaginary angular momenta, labeled by $\mu$ for the wedge and $\lambda$ for the cone, {enumerate the possible scattering waves}. The cone's scattering waves are also labeled by the real integer $m$, corresponding to the $z$-component of angular momentum. The corresponding $T$-matrices depend on the opening angle $\theta_0$.  The result for the wedge ({top-right in Fig.~\ref{elements}}) is independent of the axial wavevector $k_z$.  In addition to the usual polarizations, for the cone we needed to introduce an extra ``ghost'' field {(labelled $Gh$)} and the corresponding $T$-matrix (top-{left in Fig.~\ref{elements}}).

We will also need the matrix $\cal U$ that captures the appropriate translations and rotations between the scattering bases for each object. This matrix encodes the objects' relative positions and orientations.
{The ${\mathcal{U}}$-matrices needed for our calculations are presented in Sections~\ref{sec:wedge} and \ref{sec:cone}.}  The expression for the Casimir interaction energy,
\begin{equation}
\label{Egen}
\mathcal E={\hbar c\over 2\pi}\int_0^\infty d\kappa\,\tr\ln [1-\mathcal N]=
-{\hbar c\over 2\pi}\int_0^\infty d\kappa
\left[\tr \mathcal N+\frac{1}{2}\tr \mathcal N^2+\cdots\right],
\end{equation}
involves integration over the imaginary wave number $\kappa$, an implicit argument of the above matrices, which are combined into $\mathcal N=\mathcal UT^{\rm object}\mathcal U^{\dagger}T^{\rm plate}$ {where we considered one of the objects to be an infinite plane}.  An expansion of $\tr\ln[1-\mathcal N]$ in powers of $\mathcal N$ corresponds to multiple scatterings of quantum fluctuations of the EM field between the two objects; the trace operation sums over appropriate bases (plane waves {for example}).
This procedure can be generalized to multiple objects, with the material properties and shape of each body encoded in its $T$-matrix. Analytical results are restricted to objects for which EM scattering can be solved exactly in a multipole expansion,  a familiar problem of mathematical physics with classic applications to radar and optics.

\begin{figure}
\includegraphics[width=13cm]{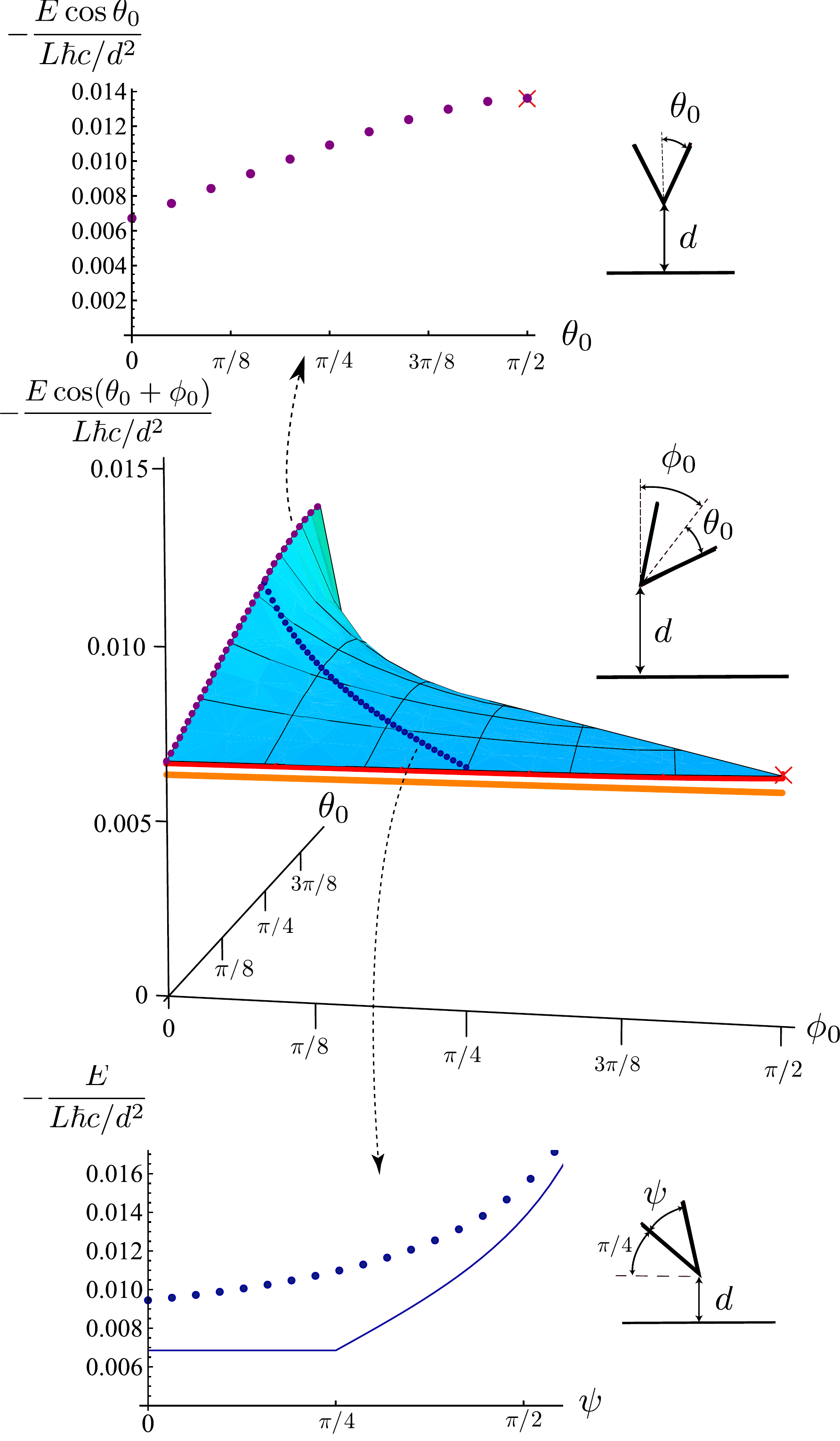}
\caption{ {The Casimir interaction energy
of a wedge at a distance $d$ above a plane, as a function of its semi-opening angle $\theta_{0}$ and tilt $\phi_0$.  The rescaled energy as a function of $\theta_{0}$ and $\phi_{0}$ is shown in the middle panel.  The symmetric case, $\phi_{0}=0$, is displayed in the top panel and the interesting case where the back side of the wedge is ``hidden'' from the plant is shown in the bottom panel.  See the text for further discussion.}
}
\label{wedge}
\end{figure}

For a perfectly reflecting wedge, translation symmetry makes it possible to decompose the EM field into two scalar components: an E-polarization field that vanishes on its surface (Dirichlet boundary condition), and an M-polarization field that has vanishing normal derivative (Neumann).  In the cylindrical coordinate system $(r,\phi,z)$, a wedge has surfaces of constant $\phi=\pm\theta_0$. Whereas for describing scattering from cylinders a natural basis is $e^{\pm im\phi} H_{m}^{(1)}(i\sqrt{\kappa^2+k_z^2} r) e^{ik_z z}$ with Bessel-$H^{(1)}$ functions indexed by $m=0,1,2,\cdots$, for a wedge we must choose $e^{\pm\mu \phi}H_{i\mu}^{(1)}(i\sqrt{\kappa^2+k_z^2} r)e^{ik_z z}$ with real $\mu\geq0$, corresponding to imaginary angular momenta that are no longer quantized (see Section~\ref{sec:wedge}).  The $T$-matrices, diagonal in $\mu$, take the simple forms indicated in Fig.~\ref{elements}.  Dimensional analysis indicates that the interaction energy of a wedge of {edge length} $L$ at a separation $d$ from a plane is $\mathcal E=-(\hbar c L/d^2) f(\theta_0,\phi_0)$, where $f(\theta_{0},\phi_{0})$ is a dimensionless function of the opening angle $\theta_0$, and inclination $\phi_0$ to the plane.  This geometry, and the corresponding function $f(\theta_0,\phi_0)$, are plotted in {the middle panel of} Fig.~\ref{wedge}.

The limit $\theta_0\,\to\,0$ corresponds to a ``knife-edge,'' which was previously studied as a limiting form of a parabolic cylinder \cite{Graham10}.  The matrix $\mathcal N$ for the wedge simplifies in this limit, enabling exact calculation of the first few terms in the expansion of $\tr\ln[1-\mathcal N]$ in Eq.~(\ref{Egen}),
\begin{align}
\label{Ewedge}
-\frac{\mathcal E}{\hbar c L/ d^2 }  = &\left[\frac{  \sec
\phi_0}{16\pi^2}\right]
+
\left[\frac{1}{192\pi^3}\right.
+\left.\frac{1}{256\pi^3}\left(\csc^3\phi_0\sec\phi_0(2
\phi_0 - \sin2 \phi_0)\right) \right]+\cdots.
\end{align}
The first square brackets corresponding to $\tr \mathcal N$ (depicted by an orange line in Fig.~\ref{wedge}) and the second to $\tr \mathcal N^2/2$; their sum (depicted by a red line) is in remarkable agreement with the full result (blue surface).  As $\phi_{0}\to \pi/2$, the knife{-}edge becomes parallel to the plane and the interaction energy diverges as it becomes proportional to the area rather than $L$. For parallel plates we know that the terms in the multiple-scattering series $\tr [{\mathcal N}^n]/n$ are proportional to $1/n^4$ \cite{Milton01}. Numerically, we find that the convergence is more rapid for $\phi_0 < \pi/2$, and that the first two terms in Eq.~(\ref{Ewedge}) are accurate to within 1\%.  { Including more than three terms in the series will not modify the curve at the level of accuracy for this figure.  Casimir's calculation for parallel plates gives an exact result at $\phi_{0}=\pi/2$, marked with an $\times$.} \

{Fig.~2 displays some of the more interesting aspects of the wedge-plane geometry.  In the middle panel the energy, rescaled by an overall factor of $\hbar c L/d^2$, and multiplied by $\cos\left(\theta_0+\phi_0\right)$, is plotted versus $\theta_{0}$ and $\phi_{0}$.}  The factor of $\cos\left(\theta_0+\phi_0\right)$ is introduced to remove the divergence as one face of the wedge becomes parallel to the plane and the energy becomes proportional to the area rather than just the length of the wedge. The blue surface is obtained by numerical evaluation of the {first three terms in the} multiple-scattering series of Eq.~(\ref{Egen}), with $\mathcal N$ constructed in terms of the plate and wedge $T$-matrices given in Fig.~\ref{elements}. The front curve ($\theta_0=0$) corresponds to a knife-edge {as previously mentioned.} {The {top} panel depicts the ``butterfly" configuration where the wedge is aligned symmetrically with respect to the normal to the plate.}  As the wings open up to a full plane at $\theta_0=\pi/2$, {the result again approaches the classic parallel plate result, again marked by a $\times$.}   {Finally, and perhaps most interestingly from a qualitative point of view,}  the {bottom} panel depicts the case where one wing is fixed at $\pi/4$, and the other opens up by $\psi=2\theta_0$. {This case displays the sensitivity of the Casimir energy to the back side of the wedge, which is hidden from the plate.  In the proximity force approximation the energy is independent of the orientation of the back side of the wedge,} and thus the PFA result (solid line) is constant until the back surface becomes visible to the plane. The correct result (dotted line) {varies continuously with the opening angle and differs from the proximity force estimate by nearly a factor of two, showing that the effects responsible for the Casimir energy are more subtle than can be captured by the PFA.}

\begin{figure}
\begin{center}
\includegraphics[width=17cm]{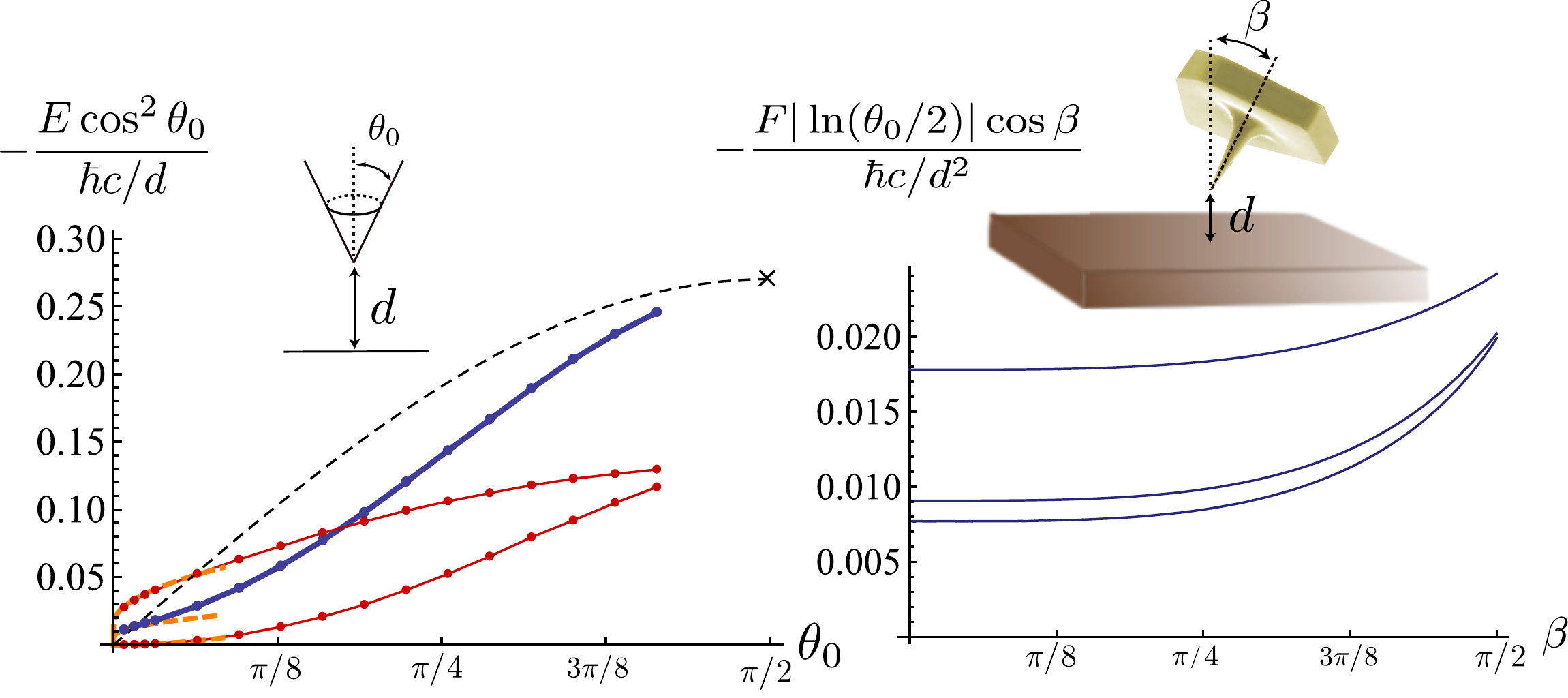}
	\caption{Casimir interaction of a cone of semi-opening angle $\theta_0$ a distance $d$ above a plane, scaled by the dimensional factor of $\hbar c/d$. The left panel corresponds to a vertical {orientation}, {with} the energy multiplied by $\cos^2\theta_0$ to remove the {divergence as the energy} becomes proportional to the area for $\theta_0=\pi/2$.  {The right panel shows the force, $F$, suitably scaled,} for a tilted, sharp cone ($\theta_0\to0$, evocative of an AFM tip) as a function of tilt angle $\beta$ and temperatures $T$=300, 80,  and 0K (top to bottom), at a separation of 1$\mu$m.  See the text for further discussion.}
 \label{Fig-cone}
 \end{center}
\end{figure}

Computations for a cone --- the surface of constant $\theta$ in {\it spherical} coordinates $(r, \theta, \phi)$ --- require a similar passage from spherical waves labeled by $(\ell,m)$ to counterparts of imaginary angular momentum, $\ell\to i\lambda-1/2$. In this case $\lambda\geq0$ is real, {while} $m$ {remains} quantized to integer values.  However, unlike the wedge case, the EM field can no longer be separated into two scalar parts{;} the more complex representation we report in Section~\ref{sec:cone} involves an additional field, similar to the ``ghost'' fields that appear in some quantum field theories.  To our knowledge, this is a novel representation of EM scattering, which should also be of use for describing reflection of ordinary EM waves from cones.  Dimensional analysis indicates that for a cone poised vertically at a distance $d$ from a plane, the interaction energy scales as $(\hbar c /d)$ times a function of its opening angle $\theta_0$. This arrangement and the resulting interaction energy are depicted in Fig.~\ref{Fig-cone} (left panel),
with the energy scaled by $\cos^2\theta_0$ to remove the divergence as the cone opens up to a plane for $\theta_0\to\pi/2$.  The PFA approximation \cite{Derjaguin56}, (depicted by the dashed line) becomes exact in this limit, but it is progressively worse as $\theta_0$ decreases from $\pi/2$.  In particular, it predicts that the energy vanishes linearly as $\theta_0\to0$, while in fact it vanishes as
\begin{equation}
\label{SharpCone}
\mathcal E\, {\sim} \,-\frac{\hbar c }{ d}  \ \frac{\ln 4-1}{16\pi}  \
\frac{1}{|\ln{\frac{\theta_0}{2}}|}~,
\end{equation}
where the logarithmic divergence is characteristic of the remnant line in this limit \cite{Emig06}.  The EM results shown {as the bold (blue) curve in Fig.~\ref{Fig-cone}} are obtained by including two terms in the series of Eq.~(\ref{Egen}), with $\mathcal N$ constructed from the plate and cone $T$-matrices in Fig.~\ref{elements},  including the additional {``ghost''}  field. We have also included the corresponding curves for scalar fields subject to Dirichlet and Neumann boundary conditions which are depicted as {fine (red) curves where the top (bottom) one corresponds to Dircihlet (Neumann) boundary condition}.
The limit of $\theta_0\to 0$  is shown in the left panel of Fig.~3 as an orange dashed line.   As the sharp tip is tilted by an angle $\beta$,  the prefactor $(\ln 4-1)/16\pi$ is replaced by $g(\beta)/\cos\beta$, where $g(\beta)$ can be computed from integrals of trigonometric functions (see Section~\ref{sec:cone}). We plot this quantity in the right panel of Fig.~\ref{Fig-cone}.

For practical applications, the above results have to be corrected for imperfect conductivity and finite temperature. The latter correction  is easily incorporated by replacing the integral in Eq.~(1) with a sum over frequencies $\kappa_n=(2\pi k_BT/\hbar c)n$. The (analytical) result for the sharp cone is reported in Section~\ref{sec:thermal}, and plotted for $d=1\mu {\rm m}$ in Fig.~\ref{Fig-cone}. Interestingly, the room temperature force is more than 100\% higher than $T=0$. In contrast, the corresponding increase for parallel plates at $d=1\mu {\rm m}$ is only about 0.1\%. This enhanced role of thermal corrections for specific geometries has been noted before \cite{Klingmueller08}, and appears essential to the design of MEM devices. Fortuitously, the increased importance of thermal corrections diminishes the effects of imperfect conductivity. While we do not compute forces for a general frequency-dependent dielectric response $\epsilon(\omega)$, we argue in Section~\ref{sec:conductivity} that for typical metals (e.g. Au or Al), even at zero temperature and for sharp cones (the cases where these corrections are largest), the corrections due to imperfect conductivity are at most around 5\% for $d=0.2\mu {\rm m}$.

Thus for separations $0.2\mu {\rm m}\,{\lesssim}\,d\,{\lesssim}\, 10\mu {\rm m}$, relevant to MEM devices and experiments, the perfect conductor results, with the important finite temperature corrections, should suffice. For example, let us consider the tip of an atomic force microscope (AFM):  At separations, $d\approx 0.2\mu {\rm m}$, where the tip may be well approximated as a metal cone, our results predict a force that is a fraction of a pico-Newton.  Such forces are at the limit of current sensitivities \cite{Castillo-Garza09}, and will likely become accessible with future improvements.  Current experiments are performed on spheres of relatively large radius $R$, where the force is greater by a factor of $(R/d)$ (a typical $R$ is 100$\mu$m).  A rounded wedge with radius of curvature $R$ falls in {an} intermediate range, with forces larger \cite{Graham10} by $\sqrt{R/d}$ than the sharp case.  The force can also be enhanced by using arrays of cones or wedges, at the cost of the difficulty of maintaining their alignment.

\section{Wedge}
\label{sec:wedge}

The Casimir energy and energy density for a wedge in isolation have been considered previously in many contexts \cite{dowker,deutsch,brevik2,brevik3,brevik4,brevik5,brevik6,brevik7,nesterenko,rezaeian,saharian,razmi}, but these works do not address the \emph{interaction} energy that leads to the Casimir force.  We first formulate the scattering theory for a single wedge and include interactions in the subsequent parts.

{For a perfect conductor that is translationally invariant in one direction, it is possible to decompose the EM field into two scalar fields that obey the Helmholtz equation with Dirichlet and Neumann boundary conditions, respectively, on the conducting surfaces. As the scattering depends only trivially on the translationally-invariant direction,
we begin by studying the scalar field in two dimensions.
In {plane} polar coordinates
the solutions are labeled by the {component of angular momentum out of the plane}, $m$,
and wave number, $k$.   The Helmholtz equation is second-order, and
therefore has two independent solutions{, which we take to be} the regular
Bessel function $J_m(kr)$, which {is finite at}
$r=0$, and the outgoing Hankel function $H^{(1)}_{m}(kr)$, which {is irregular at $r=0$, and}
\begin{figure}[h]
  \centering
  \includegraphics[scale=.3]{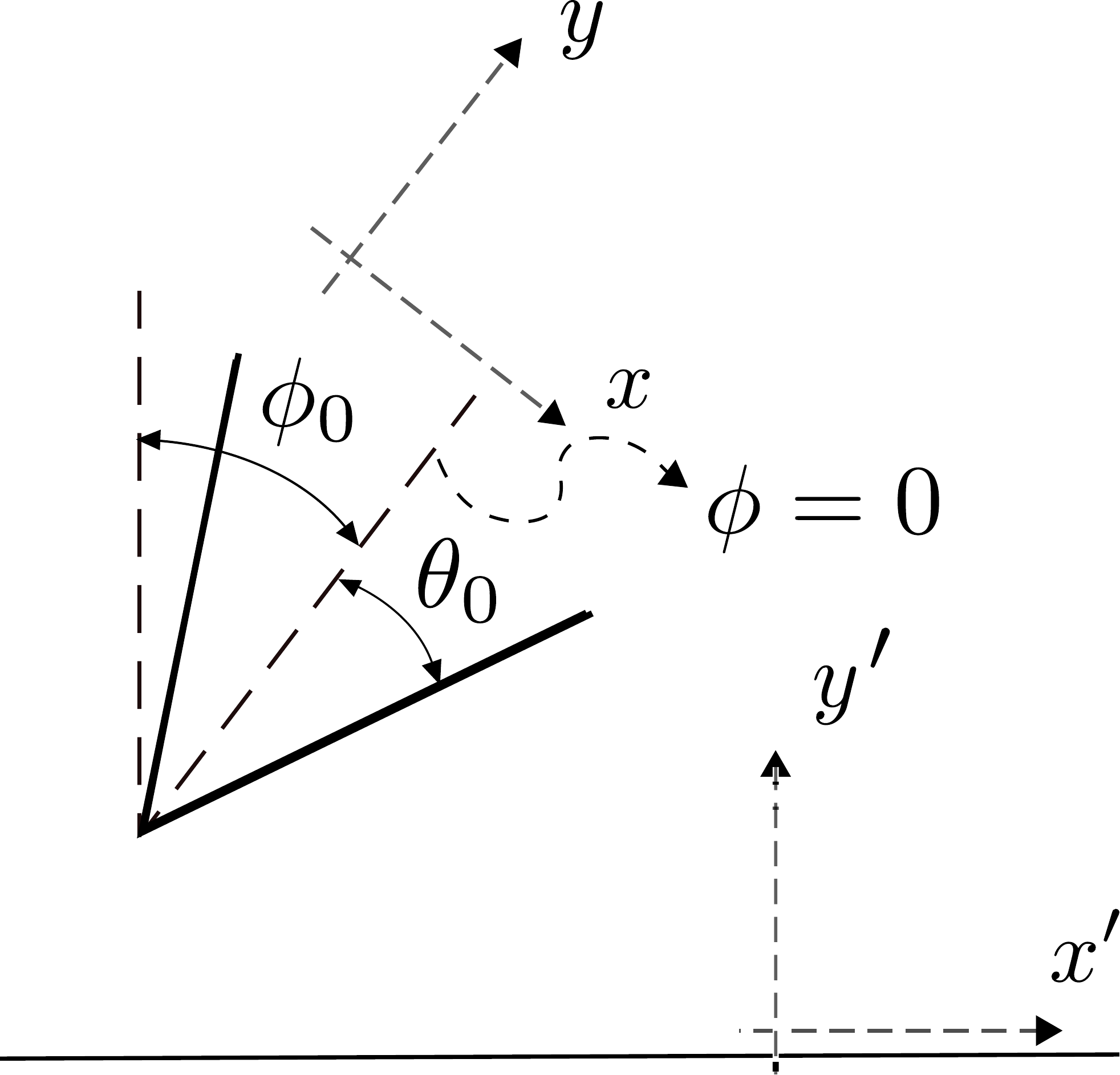} 
\caption{\label{WedgeAgainstAPlate}\small The configuration of a {tilted} wedge opposite
  an infinite plate. }
\end{figure}
obeys outgoing wave boundary conditions for $r\to \infty$.  Our
primary tool is the free Green's function in polar coordinates
for imaginary wave number $k=i\kappa$.  For applications to scattering
theory, a useful representation of this Green's function is in terms of
regular and outgoing waves \cite{Morse53},
\begin{equation}
\label{eqn:polarGreen}
G_0(r,\phi;r',\phi',i\kappa) = \frac{1}{2\pi} \sum_{m=-\infty}^\infty
e^{im(\phi-\phi')} I_m(\kappa r_<) K_m(\kappa r_>) ,
\end{equation}
in which $r_<$ ($r_>$) is the smaller (larger) of $r$ and $r'$.  For imaginary wavenumber, the regular solution becomes a modified Bessel function of the first kind $I_m(\kappa r)$, while the outgoing solution becomes a modified Bessel function of the third kind $K_m(\kappa r)$.  The former {diverges} for $r\to \infty$, while the latter {diverges} at $r=0$, but {Eq.~(\ref{eqn:polarGreen}) avoids
  these pathologies} by selecting {the regular solution for the
  smaller value of $r$ and the outgoing solution for the larger value
  of $r$.}
Along with the Green's function, we also require the expansion of
a plane wave in terms of our scattering solutions,
\begin{equation}
e^{ikr \cos \phi} = \sum_{m=-\infty}^{\infty} i^{m}
e^{i m \phi} J_m(kr)\,,
\label{eqn:JacobiAnger}
\end{equation}
where $\phi$ is the angle between $\bm{k}$ and $\bm{r}$.
The above representation is suitable for scattering from
an ordinary cylinder, for which the boundary condition is imposed {along a fixed}
value of $r$.  For the wedge, however, the boundary is instead
{defined by a constant} value of $\phi$, so we want the complementary
representation in which the discontinuity in the representation of the
Green's function is implemented through $\phi$ rather than $r$.
To create this representation, we let the polar angle $\phi$ be
defined  from $-\pi$ to $\pi${, with $+\pi$ and $-\pi$ identified}.  The points at $\phi=0$ and
{$\phi=\pm\pi$}
then serve as the
analogs of $r=0$ and $r=\infty$
in defining regular and outgoing solutions.  The symmetry axis of the
wedge is the half-line $\phi=0$ and we let $\theta_0$ be its {half-}opening angle, see Fig.~\ref{WedgeAgainstAPlate}.

We first carry out this {transformation} for the expansion of a plane wave.
From Eq.~(\ref{eqn:JacobiAnger}) with $kr=ix$, we obtain
\begin{equation} \label{Jacobi Exp}
e^{-x \cos\phi}  = {\sum_{m=0}^{\infty}}'\,\,2 I_m(x) (-1)^m \cos(m \phi ),
\end{equation}
where the prime on the sum indicates that the first term is weighted
with a factor of $1/2$.  We can represent the sum over $m$ as a
contour integral {along} $\mathcal{C}$ of an integrand with poles at
non-negative integer values of $m$.  Then the sum in Eq.
(\ref{Jacobi Exp}) becomes
\begin{equation}
e^{-x \cos\phi}  =
\int_\mathcal{C} \frac{d\nu}{2 \pi i} \frac{\pi}{\sin \nu \pi} 2
I_\nu(x)  \cos(\nu \phi ),
\end{equation}
where {the integration contour is shown in Fig.~\ref{wedgeCP}} and the factor of $1/\sin \nu \pi$ introduces poles with the
correct residues.
The functions in the integrand are analytic, so we can deform the
contour to an integral along the imaginary axis plus a semi-circle at
infinity{, which does not contribute to the integral}.
So we are left with
the integral on the imaginary axis ($\mathcal{C'}$), see Fig.~\ref{wedgeCP}. Using
\begin{equation}
K_\nu(x)=\frac{\pi}{2}\frac{I_{-\nu}(x)-I_{\nu}(x)}{\sin{\nu \pi}}{,}
\end{equation}
and the reflection symmetry of Bessel $K$ functions, we find
\begin{equation} \label{plnwave expn}
   e^{-x \cos\phi} = \frac{2}{\pi} \int_{0}^{\infty} d\lambda
   K_{i\lambda}(x) \cosh(\lambda \phi).
\end{equation}
This integral is convergent only for $|\phi|<
\pi/2$, since $K_{i\lambda} (x)$ asymptomatically goes as
$e^{-\lambda \pi/2}$ for large $\lambda$.\\
\begin{SCfigure}
\includegraphics[scale=0.27]{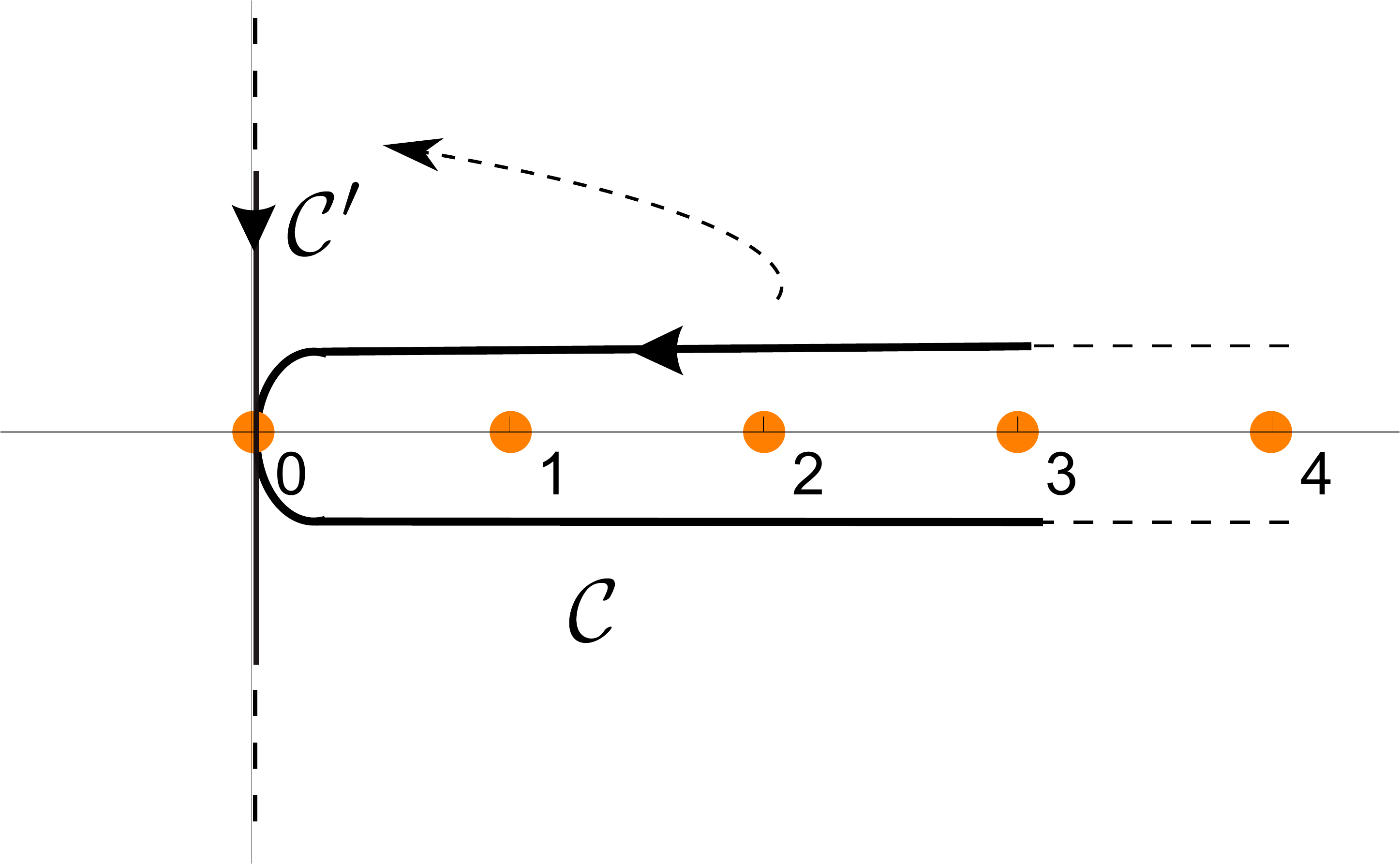}
\caption{\label{wedgeCP}\small
Analytic continuation to the imaginary axis {in the complex angular momentum plane} for the wedge.}
\end{SCfigure}
More generally we have, for arbitrary complex {angle} $a$,
\begin{equation}\label{Expsn}
e^{-x \cos(\phi-a)}   = \frac{2}{\pi} \int_{0}^{\infty} d\lambda
K_{i\lambda}(x) \cosh(\lambda (\phi-a)),
\end{equation}
which is convergent if $|\mbox{Re}(\phi-a)|<\pi/2$.  We will consider
$\phi$ and $a$ as the angles of $\bm{r}$ and $\bm{k}$ respectively.\\

The analogous representation of the Green's function can be obtained
in a similar way \cite{Oberhettinger},
\begin{align} \label{GrnFn}
    G_0(r, \phi;r' , \phi',i \kappa)
    &= \frac{1}{\pi^2} \int_{0}^{\infty} d\lambda K_{i \lambda}(\kappa
    r ) K_{i \lambda}(\kappa r' )    \cosh(\lambda(\pi-|\phi- \phi'|)) \,.
\end{align}
One can also obtain this result using the orthogonality
condition on Bessel $K$ functions with respect to their argument,
\begin{equation} \label{OrthRln}
\frac{2}{\pi^2} \int_{0}^{\infty} d\lambda \lambda
\sinh(\lambda \pi) K_{i \lambda}(\kappa r )K_{i \lambda}(\kappa r' ) =
r \delta(r-r') \,,
\end{equation}
which forms the basis for the Kontorovich-Lebedev transform \cite{KL}.\\

For scattering theory applications, it is advantageous to cast the Green's function into the same bilinear form as Eq.
(\ref{eqn:polarGreen}),
\begin{align} \label{wedge bilinear Grn fn}
G_0(r, \phi;r' , \phi', i\kappa)&=
\frac{1}{\pi^2} \int_{0}^{\infty}  d\lambda\
\Bigg[ K_{i \lambda}(\kappa
r' ) \cosh\left(\lambda(\pi- |\phi_>|)\right) K_{i \lambda}(\kappa r)
 \cosh(\lambda \phi_<) \nonumber \\
& +{\mbox{sgn}(\phi_>)} K_{i \lambda}(\kappa r' ) \sinh\left(\lambda(\pi- |\phi_>|)\right)
K_{i \lambda}(\kappa r) \sinh(\lambda \phi_<) \Bigg]\,,
\end{align}
where $\phi_<$ ($\phi_>$) is the angle with smaller (larger) absolute value ($|\phi|$, $|\phi'|$).
From this expression we can now read off the ``regular'' and
``outgoing'' solutions, which are defined as the function which are well-behaved at $\phi=0$ and $\phi=\pm
\pi$ respectively.  From Eq. (\ref{wedge bilinear Grn fn}),
we have {
\begin{align}
      \Phi^{{\rm reg}, c}(r, \phi)&= K_{i \lambda}(\kappa r) \cosh(\lambda
\phi), \nonumber\\
  \Phi^{{\rm out}, c}(r', \phi')&= K_{i \lambda}(\kappa r')
\cosh(\lambda (\pi-|\phi'|)), \nonumber \\
      \Phi^{{\rm reg}, s}(r, \phi)&= K_{i \lambda}(\kappa r) \sinh(\lambda
\phi), \nonumber\\
 \Phi^{{\rm out}, s}(r', \phi')&= K_{i \lambda}(\kappa r')
\sinh(\lambda (\pi-|\phi'|)) \mbox{sgn}(\phi')  \,,
\end{align}
}
\begin{figure}[h]
  \centering
  \includegraphics[scale=.22]{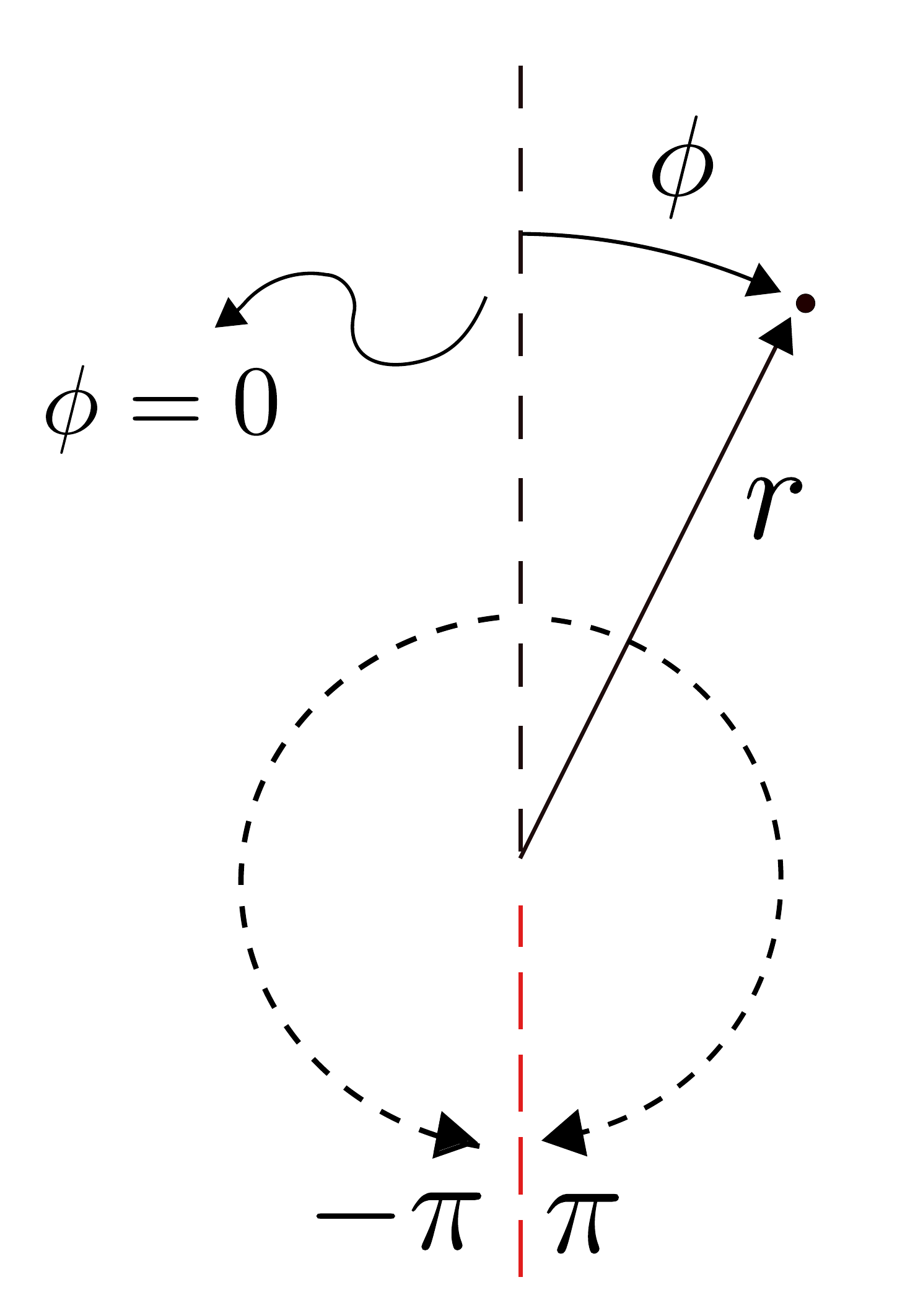}
\caption{\small {Plane polar coordinates appropriate for the wedge}.  The discontinuity at {$\phi=\pm\pi$}  constrains the definition of regular and outgoing functions.}
\end{figure}
where the superscripts $c$ and $s$ stand for symmetric and antisymmetric
wavefunctions in $\phi$ respectively. Note that the regular functions are {irregular at $\phi=\pm\pi$}.
Also the outgoing functions are {not regular at $\phi=0$.}
We also define the normalization coefficients
\begin{equation}
   C^{c/s}_{\lambda}=\frac{1}{\pi^2} \, ,
\end{equation}
which becomes important upon changing to a different basis, see Eq. \eqref{eqn:wedgeT}.

Dirichlet and Neumann boundary conditions on the wedge
are satisfied by an appropriate linear combination of the
regular and outgoing functions, from which we can read off the
$T$-matrix elements,
\begin{align} \label{T-wedge}
   & T^{c}_{ D\lambda}= -\frac{\cosh(\lambda \theta_0)}{\cosh(\lambda
   (\pi-\theta_0))}, \hskip .3in  T^{s}_{ D\lambda}=-\frac{\sinh(\lambda
   \theta_0)}{\sinh(\lambda (\pi-\theta_0))} \,,
   \nonumber \\
   & T^{c}_{ N\lambda}=\frac{\sinh(\lambda \theta_0)}{\sinh(\lambda
   (\pi-\theta_0))}, \hskip .3in  T^{s}_{ N\lambda}= \frac{\cosh(\lambda
   \theta_0)}{\cosh(\lambda (\pi-\theta_0))} \,,
\end{align}
where $\theta_0$ is the half-opening angle of the wedge.   Here D and N
stand for Dirichlet and Neumann boundary conditions respectively{\footnote{{The Dirichlet and Neumann boundary conditions in the case of scalar field correspond to electric ($E$) and magnetic ($M$) modes for electromagnetism, respectively. $(D, N)$ and $(c,s)$ are referred to as $(E,M)$ and $(+,-)$ in Fig~1.}}}.\\

Now let us return to a wedge with half-opening-angle $\theta_0$
opposite an infinite plate.
We take $\phi =0$ to be the symmetry
axis of the wedge.  To begin with, we consider the case where this
axis is perpendicular to the plane.  Then the $y'$ axis, defined as
the perpendicular {axis} to the plane, will be parallel to $y$, the axis along the $\phi=0$ line.
The wavevector, $\bm{k}=(k_{x'}, k_{y'})$, satisfies the {on-shell condition}, hence
\begin{equation}
k_{x'}^2+k_{y'}^2=-\kappa^2.
\end{equation}
Of course, $k_{x'}$ and $k_{y'}$ cannot both  be real. In fact, $k_{x'}$, the component of the wavevector parallel to plane, should be real while $k_{y'}=i\sqrt{\kappa^2+k_{x'}^2}$ is imaginary. So the angle $a$ as defined by
\begin{equation}
\bm{k}= (k_{x'}, k_{y'})=(k_{x'}, i(\kappa^2+k_{x'}^2)^{1/2})=(i\kappa
\sin a, i \kappa \cos a)
\end{equation}
is imaginary, and whose range is
$(-i\infty, +i\infty)$.
Note that Eq. (\ref{Expsn})
is convergent for all values of $\phi\in (-\pi/2,\pi/2)$ (of which
the wedge is a subset) and can be more conveniently written as
\begin{equation}
e^{i\bm{k} \cdot \bm{r}} =e^{-\kappa r \cos(\phi-a)}   = \frac{2}{\pi}
 \int_{0}^{\infty} d\lambda \left(K_{i\lambda} (\kappa r)
 \cosh(\lambda \phi) \cosh(\lambda a)+ K_{i\lambda} (\kappa
 r)\sinh(\lambda \phi)\sinh(\lambda a)\right).
\end{equation}
The conversion matrix elements \cite{Rahi09} 
between the wedge and
the plane scattering waves  are then given by\footnote{The {$x$-index on $k_{x}$ is unprimed because the conversion matrices are defined with respect to the $(x,z)$} axes of the wedge.}
\begin{equation}
  D^c_{\lambda,k_x}=\frac{2}{\pi}\cosh(\lambda a), \hskip.4in
  D^s_{\lambda,k_x}=\frac{2}{\pi}\sinh(\lambda a) \,.
\end{equation}
If the plane is tilted by an angle $\phi_0$, we have instead
 \begin{equation} \label{Conv-Wedge}
   D^c_{\lambda,k_x}=\frac{2}{\pi}\cosh(\lambda (a-\phi_0)),
  \hskip.4in   D^s_{\lambda,k_x}=\frac{2}{\pi}\sinh(\lambda (a-\phi_0)) \,,
 \end{equation}
where the range of $a$ is still given by $(-i\infty, +i\infty)$.

So far, we considered the two-dimensional problem and ignored the axis
$z$ along which the geometry is translationally invariant. Now we
{rename the two dimensional wave number in the argument of the  Bessel
  functions, which we previously called $\kappa$, to be $p$, and use
  $\kappa$ to denote the (imaginary) wave number ($\omega = i\kappa
  c$) of the three-dimensional problem.}  So, $p=\sqrt{\kappa^2 +
  k_z^2}$ with $k_z$ being the wave number in the transverse
direction. Because of the translational symmetry, we can make a change
of variable to eliminate $\kappa$ and recast the three-dimensional
problem in a two-dimensional basis; the Casimir energy takes the form \cite{Rahi09, Emig07}
\begin{equation}
\label{CasimirE-2d}
\mathcal{E} = \frac{\hbar c}{2 \pi} \frac{L_z}{2} \int_{0}^{\infty} dp\, p\,\mbox{tr}\ln\left(\mathcal I- T_1 \mathcal{U}_{12}T_2 \mathcal{U}_{21}\right).
\end{equation}
 Here $L_z$ is the length of the objects in the transverse direction, $\mathcal{U}_{12}$ and $\mathcal{U}_{21}$
denote the translation matrices between
objects and $T_1$ and $T_2$ are the objects' individual $T$-matrices. Note that the ``tr'' in the last equation is integrating over all quantum numbers in the two dimensions (the third direction {being included in $p$}).
Using the $T$-matrix of the wedge from Eq. (\ref{T-wedge}) and the
conversion matrix of Eq. (\ref{Conv-Wedge}), we can express
$T$-matrix of the wedge in the plane-wave basis as
\begin{align}
\label{eqn:wedgeT}
 T_{k_x, k_x'}
   &=  \sum_{A=c,s}\int_{0}^{\infty} d\lambda \int_{0}^{\infty}
   d\lambda' \frac{C_{k_x}}{C^A_\lambda} D^{\dagger A}_{\lambda, k_x}
   T^{A}_{\lambda, \lambda'} D^{A}_{\lambda', k_x'}\,,
\end{align}
where $C_{k_x}=\frac{1}{2\sqrt{k_x^2+p^2}}$ is the normalization coefficient in planar-wave basis as defined in Ref.~\cite{Rahi09}. 
We define $a=i\alpha$ so that $\alpha$ is real, and convert the $k_x$
index in Eq. (\ref{eqn:wedgeT}) to an $\alpha$ index.  We then
take $\alpha$ as the outer index in Eq. (\ref{CasimirE-2d}), with
the $\lambda$ index contracted in the matrix multiplication, so that the
energy takes the form
\begin{align}
\label{eq:E_from_N}
  \mathcal{E} =\frac{\hbar c}{2 \pi}\frac{L_z}{2} \int_{0}^{\infty} dp \, p \,
  \mbox{tr}\ln\left(\mathcal{I}_{\alpha,\alpha'} - \mathcal{N}_{\alpha,
  \alpha'}\right) \,,
\end{align}
where
\begin{align}
  \mathcal{N}{^{D}}_{\alpha, \alpha'}= e^{-p d(\cosh\alpha+\cosh\alpha')}
  \frac{1}{\pi}\int_{0}^{\infty} d\lambda \Big(&
  \frac{\cosh(\lambda(i \alpha+\phi_0))\cosh(\lambda(i\alpha'-\phi_0))\cosh(\lambda \theta_0)}{\cosh(\lambda (\pi-\theta_0))} \nonumber \\
  & -\frac{\sinh(\lambda(i \alpha+\phi_0))\sinh(\lambda(i \alpha'-\phi_0))\sinh(\lambda \theta_0)}{\sinh(\lambda (\pi-\theta_0))}\Big),
\end{align}
for the Dirichlet case, and
\begin{align}
  \mathcal{N}{^{N}}_{\alpha, \alpha'}= e^{-p d(\cosh\alpha+\cosh\alpha')}
  \frac{1}{\pi}\int_{0}^{\infty} d\lambda \Big(&-
  \frac{\sinh(\lambda(i \alpha+\phi_0))\sinh(\lambda(i \alpha'-\phi_0))\cosh(\lambda \theta_0)}{\cosh(\lambda (\pi-\theta_0))}\nonumber \\
  &+\frac{\cosh(\lambda(i \alpha+\phi_0))\cosh(\lambda(i\alpha'-\phi_0))\sinh(\lambda \theta_0)}{\sinh(\lambda (\pi-\theta_0))}
 \Big),
\end{align}
for the Neumann case, in which $d$ is the distance between the plane and
the {edge of the} wedge and the trace is defined as tr\,$f(\alpha,
\alpha')=\int_{-\infty}^{\infty}d\alpha f(\alpha, \alpha)$.

{The integral over $\lambda$}
can be evaluated exactly, but we first consider a special
case.  In the limit that the wedge becomes a {knife edge},{ \it i.e.}, $\theta_0=0$,
the second term in the $\mathcal{N}$-matrix vanishes and we find
\begin{equation}
  \left. \mathcal{N}{^{D/N}}_{\alpha, \alpha'}\right|_{{\theta_0=0}} =   \frac{1}{4 \pi} e^{-p d(\cosh\alpha+\cosh\alpha')} \left( \pm \sec(i(\alpha+\alpha')/2)+\sec(i(\alpha-\alpha')/2+\phi_0)\right) \, .
\end{equation}
We note that we can also compute the tr\,$\ln$ with $\lambda$ rather
than $\alpha$ as the outer index.  The $\mathcal{N}$-matrix in this
basis is given by
\begin{equation}
\left.\mathcal{N}{^{D/N}}_{\lambda, \lambda'}\right|_{{\theta_0=0}} =   \frac{1}{\pi\cosh(\lambda
\pi)} \left( \pm K_{i(\lambda+\lambda')}(2p d)
\cosh((\lambda-\lambda')\phi_0) + K_{i(\lambda-\lambda')}(2p
d)\cosh((\lambda+\lambda')\phi_0)\right)\,,
\end{equation}
and the trace is now defined by tr\,$f(\lambda,
\lambda')=\int_{0}^{\infty}d\lambda f(\lambda, \lambda)$.

In either description, the indices of the matrix $\mathcal{N}$ are
continuous, not discrete.  In matrix multiplication, a continuous
index is {defined simply}
by replacing the sums by integrals.
Computing the tr\,$\ln$ (or, equivalently, $\ln \det$) of a
continuous matrix can also be done by discretization of the continuous
index, but for our purposes another approach will be more efficient:
We will expand $\mbox{tr}\ln(1-\mathcal{N})$ as a power series in
$\mathcal{N}$ and show that we can achieve a sufficient precision by
keeping only the first few terms.

The results for a tilted {knife edge} opposite an infinite plate are
plotted in Fig.~\subref{DirNeuTilted}\!\!. 
This plot is in agreement with the results obtained for the knife edge
as the extreme limit of a parabolic cylinder \cite{Graham10}. 
They have been computed by considering the first three
terms in the expansion of the tr\,$\ln$ {in Eq.~\eqref{eq:E_from_N}}. The rapid
convergence in powers of $\mathcal{N}$ is shown in
Fig.~\subref{DirTilted}\!\!, using the Dirichlet case as an example.
\begin{figure}[ht]
\centering
\subfigure[Dirchlet, Neumann, and EM boundary conditions.]{
\includegraphics[scale=.77]{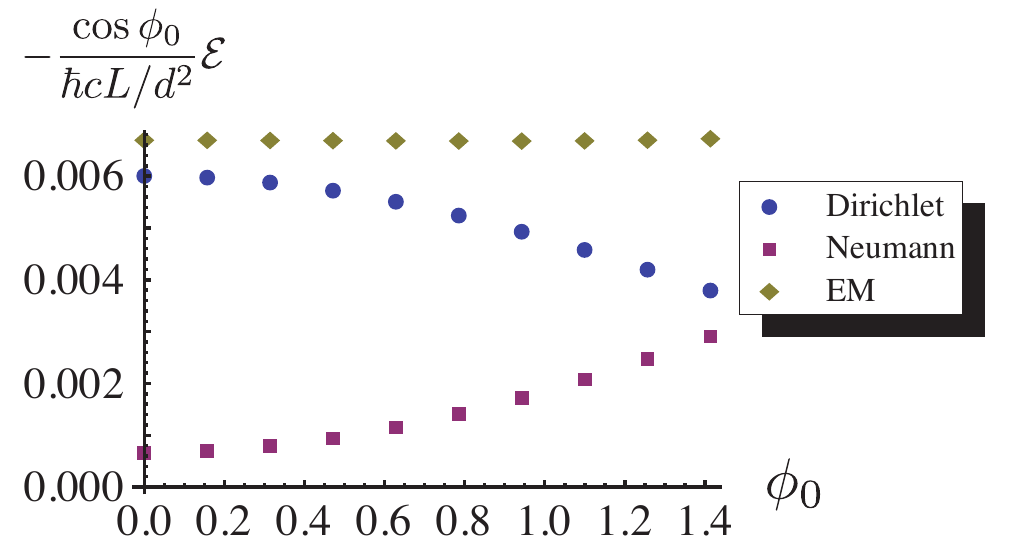}
\label{DirNeuTilted}
}\qquad
\subfigure[Convergence of
multiple-reflection expansion for Dirichlet boundary condition.]{
\includegraphics[scale=.71]{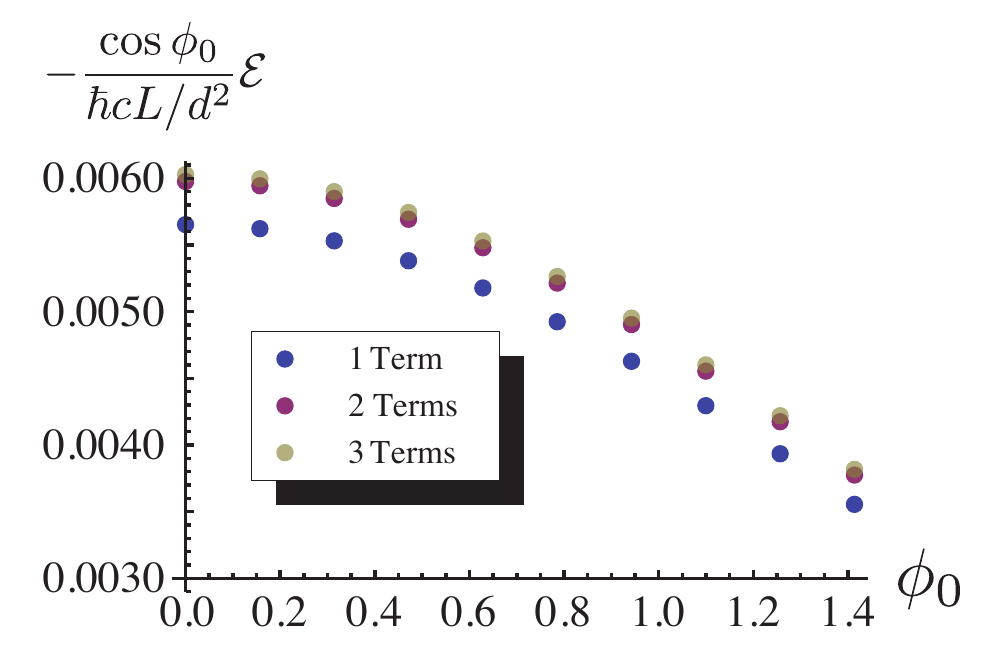}
\label{DirTilted}
}
\caption{\small Tilted {knife edge.}}
\end{figure}
For a {knife edge}
($\theta_0=0$), we can exactly compute
the first two terms in the expansion of tr\,$\ln$ formula ({\it i.e.,} the expansion in multiple reflections). First, we will report results separately for
the Dirichlet and Neumann cases {(denoted by D and N respectively)}. The electromagnetic result is
the sum of the two.  We obtain {results} that are accurate to {about} one percent by keeping the first two terms in the expansion.
Using dimensional analysis, the energy is proportional to
$\hbar c L_z/d^2$, so we calculate the scaled energy
\begin{align}
  -{\frac{\mathcal{E}^{ \substack{ \rm knife\\  \rm edge}}_{D/N}}{\hbar c  L_z/d^{2}}}= & \frac{1}{16\pi^2}\left(\pm
  \frac{\pi}{8}+\frac{1}{2} \sec \phi_0 \right) \nonumber \\
  & +\frac{1}{32\pi^3} \left[ \pm\left(\frac{\pi}{8}\right)^2
  \sec^2(\phi_0/2)+\frac{1}{12}+\frac{1}{16} \left( \csc^3\phi_0
  \sec\phi_0 (2\phi_0 -\sin2\phi_0)\right) \right] {+ \cdots} \, ,
\end{align}
where the first line is obtained by evaluating tr\,$\mathcal{N}$ and
{the expression in the }second line by evaluating $\frac{1}{2}\mbox{tr}\mathcal{N}^2$.  {The dots indicate corrections from higher order reflections.} It
is interesting to study these results in a few different limits. When
the {knife edge}
is perpendicular to the infinite plate ($\phi_0=0$),
the previous equation reduces to
\begin{equation}
  -{\left.\frac{\mathcal{E}^{ \substack{ \rm knife\\  \rm edge}}_{D/N}}{\hbar c  L_z/d^{2}}\right|_{\phi_{0}=0}}=
  \pm\frac{17}{2048\pi}+\frac{1}{32\pi^2}+\frac{1}{192\pi^3} {+\cdots} \, ,
\end{equation}
a result that heretofore was known only numerically \cite{Graham10}.  In contrast, the proximity force approximation (PFA) {vanishes} for this
  configuration.  On the other hand, in the limit where the {knife edge}
  is almost parallel to the infinite plate ($\phi_0=\pi/2-\epsilon$), we have
  \begin{align}
    -{\left.\frac{\mathcal{E}^{ \substack{ \rm knife\\  \rm edge}}_{D/N}}{\hbar c  L_z/d^{2}}\right|_{\phi_{0}=(\pi/2-\epsilon)}}=
    & \frac{1}{32\pi^2} {\left(1+\frac{1}{16}\right) \frac{1}{\epsilon}}+
    \left(\pm\frac{9}{1024\pi}
      -\frac{1}{192 \pi^3}\right)+ O(\epsilon){+\cdots} \,.
    \label{eqn:01}
  \end{align}
  The first term diverges as $\epsilon \to 0$ because it gives the
  contribution proportional to the area that arises as the plates become
  parallel.  The second term then represents the leading edge correction
  which is in reasonable agreement with Ref.~\cite{Graham10} 
  for Dirichlet and Neumann boundary conditions. However, the
  electromagnetic result that sums the two varies very slowly (see
  Fig.~\subref{DirNeuTilted}\!\!) and thus the edge term (which is sensitive to
  the relative accuracy of nearby points) {deviates from the result in Ref.~\cite{Graham10}. 
We expect convergence to the exact result by including higher-order terms in multiple reflections.}

As noted above, the {expansion of the logarithm in} $\mathcal{N}$ corresponds to a
  multiple reflection expansion.  This identification forms the basis
  for the optical approximation to the Casimir energy
  \cite{Scardicchio:2004fy}.  For parallel plates, the terms in this series
  fall in magnitude like $1/n^4$, where $n$ is the number of reflections
  (back and forth) between the objects.  We can check this result
  explicitly for the area term.  {Eq.~(\ref{eqn:01}) displays the first two terms, $1 +\frac{1}{16}$, in the $1/n^{4}$ series which sums to $\zeta(4)=\pi^{4}/90$.  The contribution of the first two terms captures more than 98\% of the exact result.}
  For a wedge (or a {knife edge}) away from $\phi_0=\pi/2$,
  we find numerically that higher reflections fall off even
  more rapidly than $1/n^4$.  For as many as six reflections,
  for a given $\phi_0 < \pi/2$, the fall-off is a good fit to
  $1/n^{4+\delta(\phi_0)}$ where $\delta(\phi_0)${ is a positive function of the angle}.

  Finally, the electromagnetic Casimir energy including the first two reflections, is
  \begin{align}
    -{ \frac{\mathcal{E}^{ \substack{ \rm knife\\  \rm edge}}_{\rm EM}}{\hbar c  L_z/d^{2}}} = & \frac{1}{16\pi^2}  \sec
    \phi_0
    +  \frac{1}{256\pi^3} \left(
      \frac{4}{3}+ \csc^3\phi_0\sec\phi_0(2 \phi_0 -
      \sin2 \phi_0) \right) {+ \cdots} \, .
     \end{align}

  When the opening angle of the wedge is nonzero, {the} $\mathcal{N}$-matrix
  takes a more complicated form,
  \begin{align}
    \mathcal{N}{^{D/N}}_{\alpha, \alpha'}= \frac{1}{8(\pi-\theta_0)} e^{-p
      d(\cosh\alpha+\cosh\alpha')} \times \Bigg( & \pm
    \sec\left(\frac{\pi(\theta_0+i\alpha+i\alpha')}{2(\pi-\theta_0)}\right)+
    \sec\left(\frac{\pi(\theta_0+i\alpha-i\alpha'+2\phi_0)}{2(\pi-\theta_0)}\right)
    \nonumber \\
    &+  \sec\left(\frac{\pi(\theta_0-i\alpha+i\alpha'-2
        \phi_0)}{2(\pi-\theta_0)}\right)\pm
    \sec\left(\frac{\pi(\theta_0-i\alpha-i\alpha')}{2(\pi-\theta_0)}\right)
    \nonumber \\
    &\pm \cot\left(\frac{\pi(\pi+i
        \alpha+i\alpha')}{2(\pi-\theta_0)}\right)-\cot\left(\frac{\pi(\pi+i
        \alpha-i\alpha'+2 \phi_0)}{2(\pi-\theta_0)}\right) \nonumber \\
    & -\cot\left(\frac{\pi(\pi-i
        \alpha+i\alpha'-2\phi_0)}{2(\pi-\theta_0)}\right) \pm
    \cot\left(\frac{\pi(\pi-i
        \alpha-i\alpha')}{2(\pi-\theta_0)}\right)\Bigg),
    \label{withopening}
  \end{align}
  or, in the $\lambda$ basis,
  \begin{align}
    \mathcal{N}{^{D/N}}_{\lambda, \lambda'}= \frac{2}{\pi} \Big(&\pm
    K_{i(\lambda+\lambda')}(2p d) \ \cosh((\lambda-\lambda')\phi_0)
    \frac{\sinh(\lambda(\pi- 2 \theta_0))}{\sinh(2 \lambda (\pi-\theta_0))}
    \nonumber \\
    &+ K_{i(\lambda-\lambda')}(2p d)\cosh((\lambda+\lambda')\phi_0) \
    \frac{\sinh(\lambda\pi)}{\sinh(2 \lambda (\pi-\theta_0))}\Big).
  \end{align}
  \begin{figure}[t]
    \centering
    \subfigure[Dirchlet, Neumann and EM boundary conditions.]{
      \includegraphics[scale=.76]{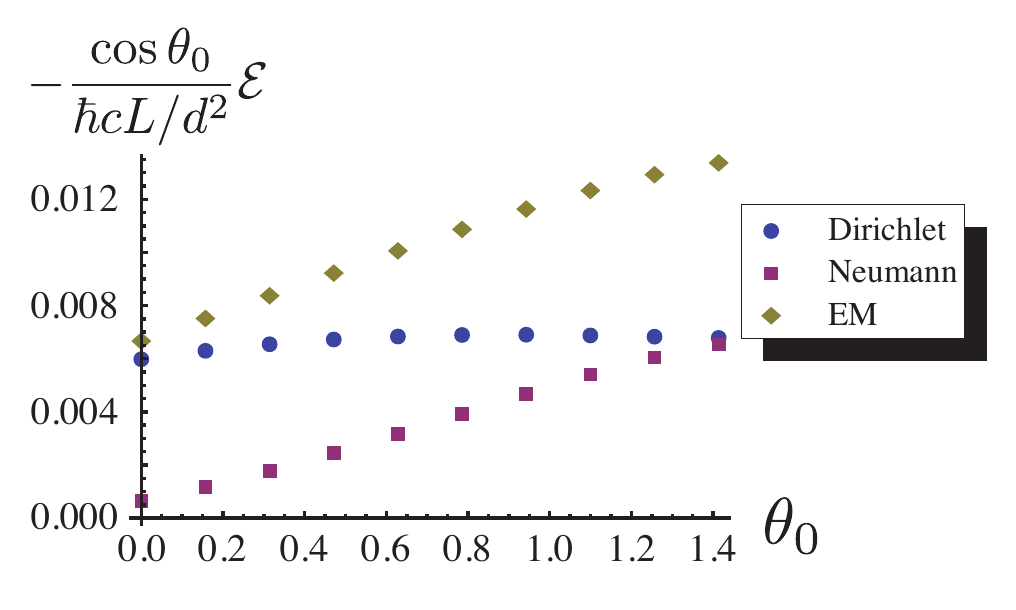}
      \label{DirNeuElecNotTilted}
    }\qquad
    \subfigure[Convergence of the
    reflection expansion for Dirichlet conditions.]{
      \includegraphics[scale=.72]{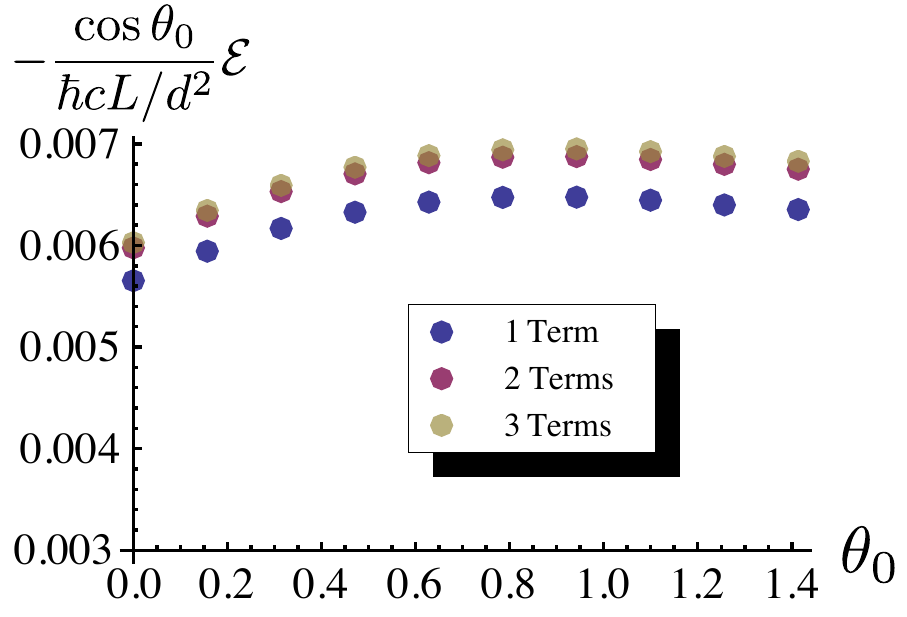}
      \label{DirNotTilted}
    }
    \caption{\small {Vertical} wedge.}
  \end{figure}

  \begin{figure}[b]
  \begin{center}
      \subfigure[$\phi_0+\theta_0=\pi/8$]{
        \includegraphics[scale=.41]{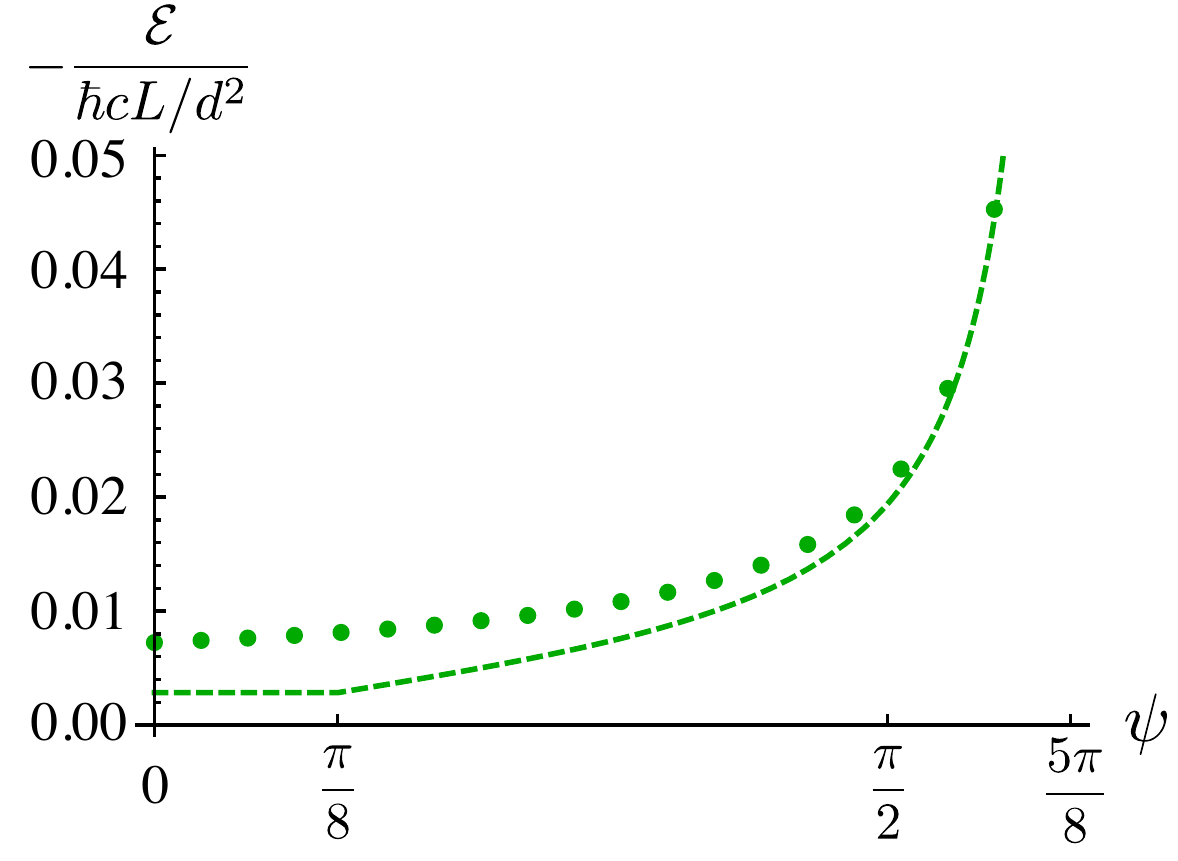}
        \label{1}
      }\qquad
      \subfigure[$\phi_0+\theta_0=\pi/4$]{
        \includegraphics[scale=.41]{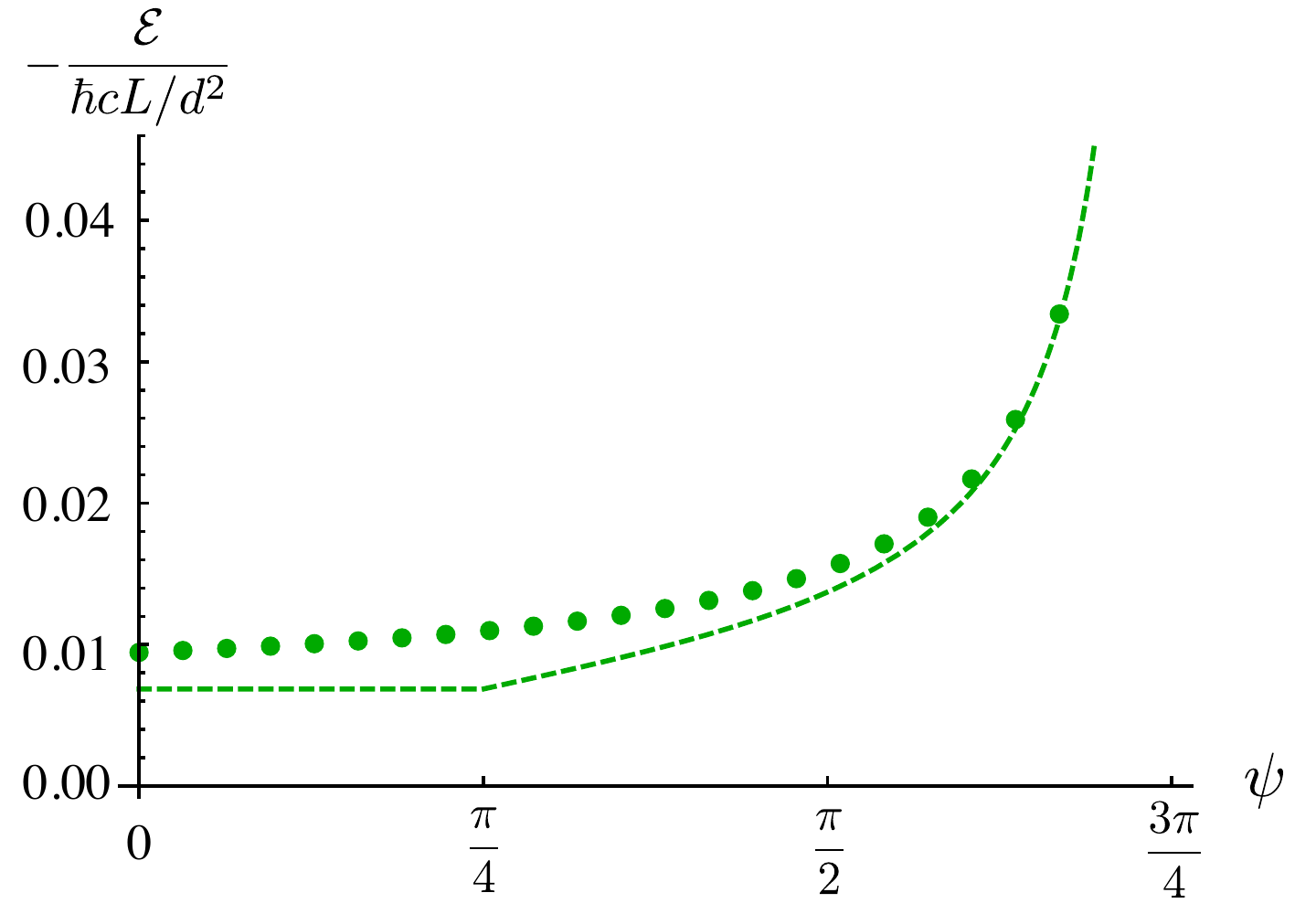}
        \label{2}
      }
      \\
      \subfigure[$\phi_0+\theta_0=3\pi/8$]{
        \includegraphics[scale=.41]{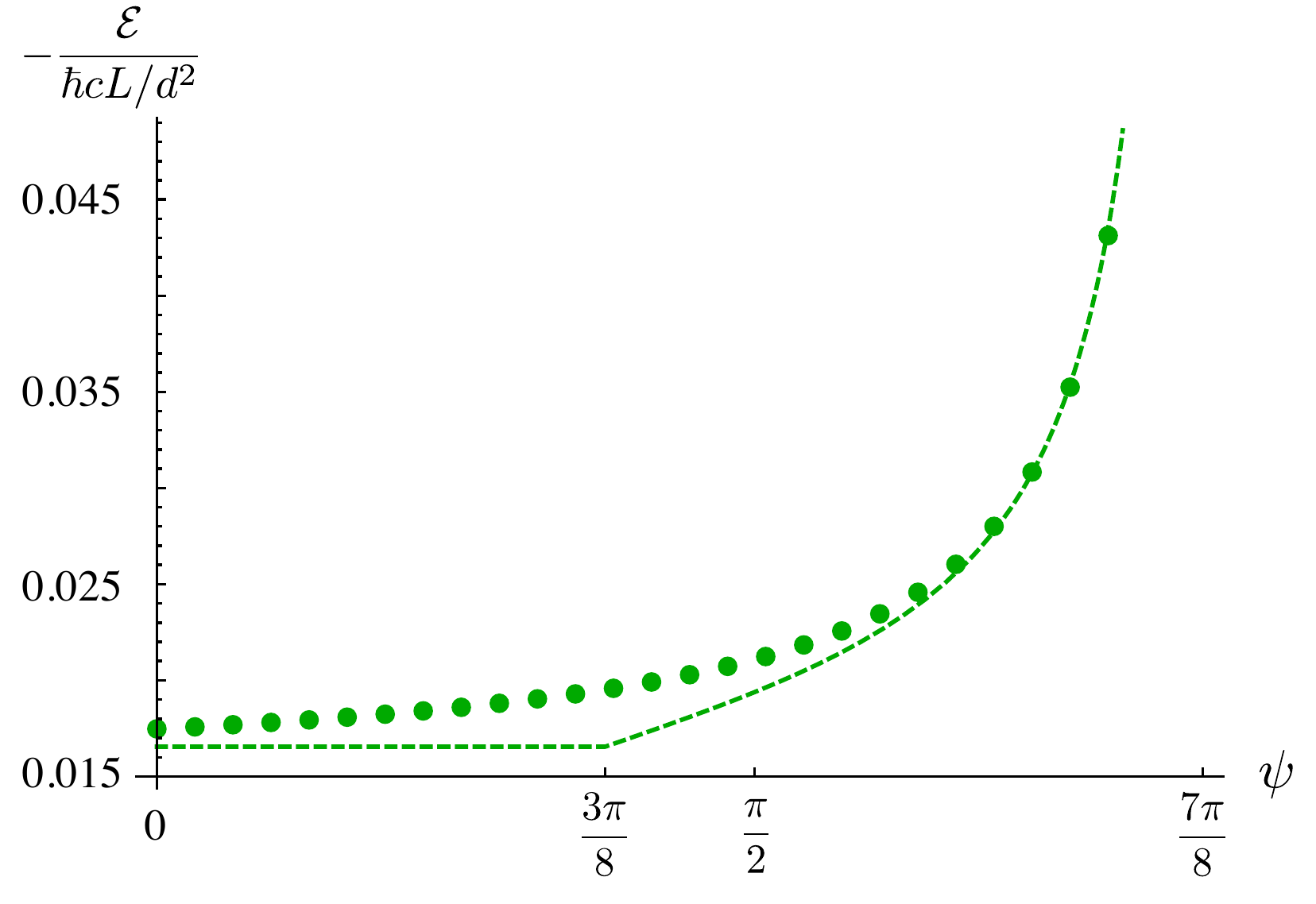}
        \label{3}
      }
      \subfigure[Definition of the angles.]
      {
      \def\svgwidth{3cm}
      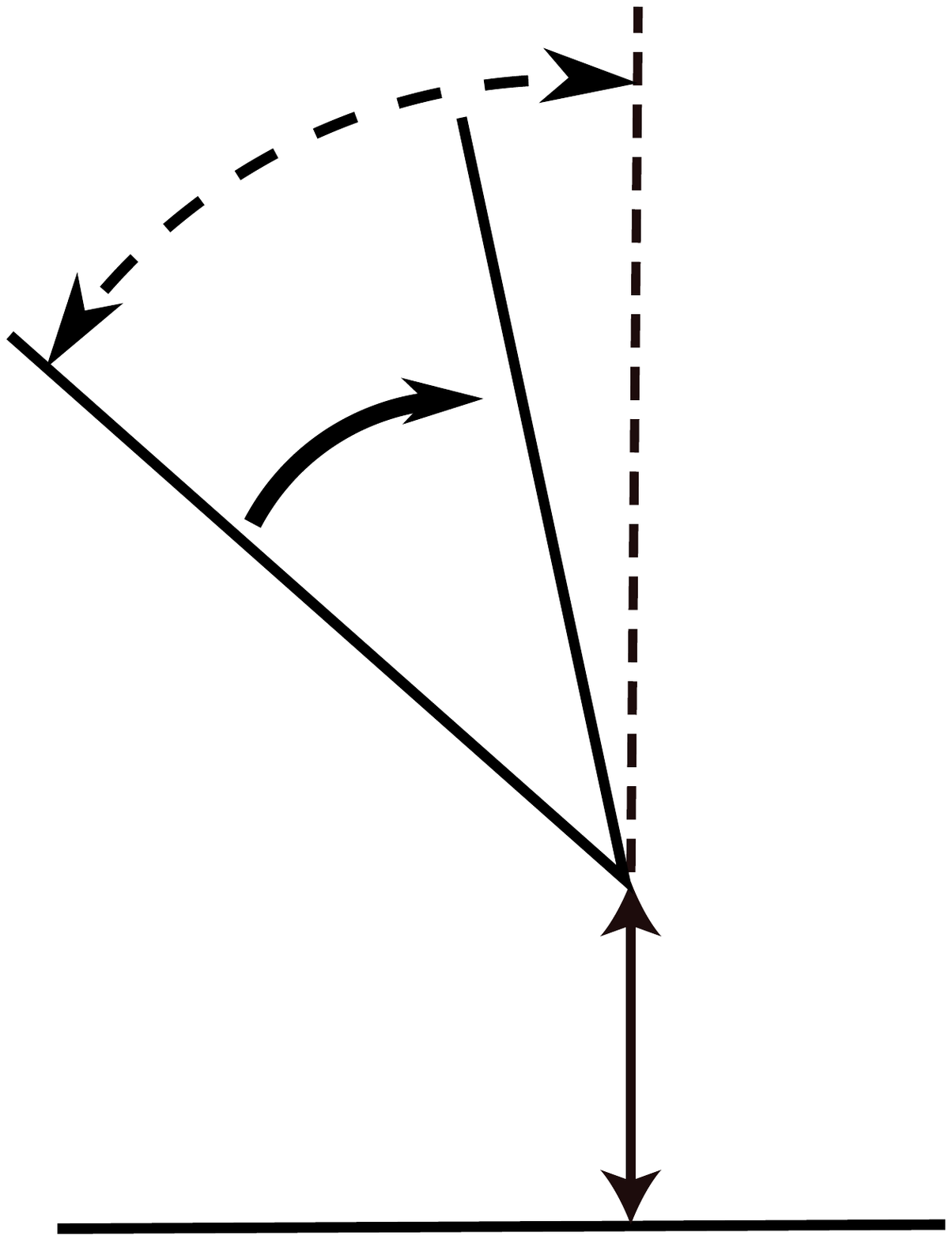
      \label{1.5}
      }
      \caption{\small {Casimir energy of a wedge with one face at a fixed angle $\phi_0+\theta_0$ as a function of the full opening angle $\psi=2\theta_0$. The} PFA prediction{, in dashed lines, is compared to the exact calculation, indicated by solid circles, where 3 multiple reflections have been taken into account.} The PFA does not change until the top face becomes visible to the plane {($\psi=2\theta_0>\phi_0+\theta_0$)}.}
  \end{center}
  \end{figure}
  The energies for a wedge of opening angle $\theta_0$ {positioned
    vertically above the plane, {\it i.e.\/}, with} tilt angle
  $\phi_0=0$, are plotted in Fig.~\subref{DirNeuElecNotTilted}\!\!, including three terms of the reflection expansion. The
  convergence in multiple reflections is shown for the Dirichlet
  boundary condition in Fig.~\subref{DirNotTilted}\!\!.

 As in the case of the knife edge, we can find analytical results for the wedge (with non-zero opening angle). {Only including the first term in multiple reflections, the electrodynamic Casimir energy is given by}
  \begin{equation}
   -\frac{\mathcal{E}^{{\rm wedge}}_{{\rm EM}}}{\hbar c L_z/ d^2} =  \frac{1}{16\pi^2}
    \frac{1}{{2(1 -\theta_0/\pi) }}\left({\tan \frac{\pi (\pi -2 \phi_0
    )}{4 (\pi -\theta_0 )}+\tan\frac{ \pi (\pi +2 \phi_0 )}{4 (\pi -\theta_0
    )}}\right) {+ \cdots} \,,
\label{withopeningexpansion}
\end{equation}
{where the dots indicate higher reflections.}
For  Dirichlet and Neumann boundary conditions, the expression for the
energy is more complicated because the terms of opposite sign in
Eq. (\ref{withopening}) do not cancel.  These terms do not
contribute to the torque, however, which in both cases is simply
one half of the derivative of Eq. (\ref{withopeningexpansion})
with respect to $\phi{_0}$. Hence, to this order, the torque is the same for Dirichlet and Neumann boundary conditions.

The geometry of the wedge provides an interesting example to examine the PFA prediction. Within PFA, the energy is computed by integrating over the surfaces facing each other, so that the energy {remains} constant until the back surface {of the wedge} becomes visible to the plane. {Our result of Eq.~\eqref{withopeningexpansion}}, on the other hand, depends smoothly on the angles. {To demonstrate the failure of the PFA, we} consider a wedge with one face fixed, at an angle $\pi/2 -(\phi_0+\theta_0)$ with respect to the plane, while the other face opens up, {as shown in Fig.~\subref{1.5}}\!\!. The energy
as a function of the (full) opening angle $\psi=2 \theta_0$ is shown in Fig.~\subref{1}\!\!-(c).

\section{Cone}
\label{sec:cone}
\subsection{Scalar field}
\label{subsec:scalarcone}
We now apply the same techniques to the case of {a conical perfect conductor opposite a conducting plate.}  We
start with spherical coordinates {with the origin at the tip of the cone and the $z$-axis aligned to the cone's symmetry axis}.

The expansion of a plane wave in terms
of spherical Bessel functions with imaginary wavenumber reads
\begin{equation}
  e^{-\kappa r \cos\Theta}= \sum_{n=0}^{\infty} (2n +1) (-1)^n
  P_n(\cos \Theta) i_n(\kappa r)\, ,
\end{equation}
\begin{figure}[h]
  \begin{center}
  \includegraphics[scale=.2]{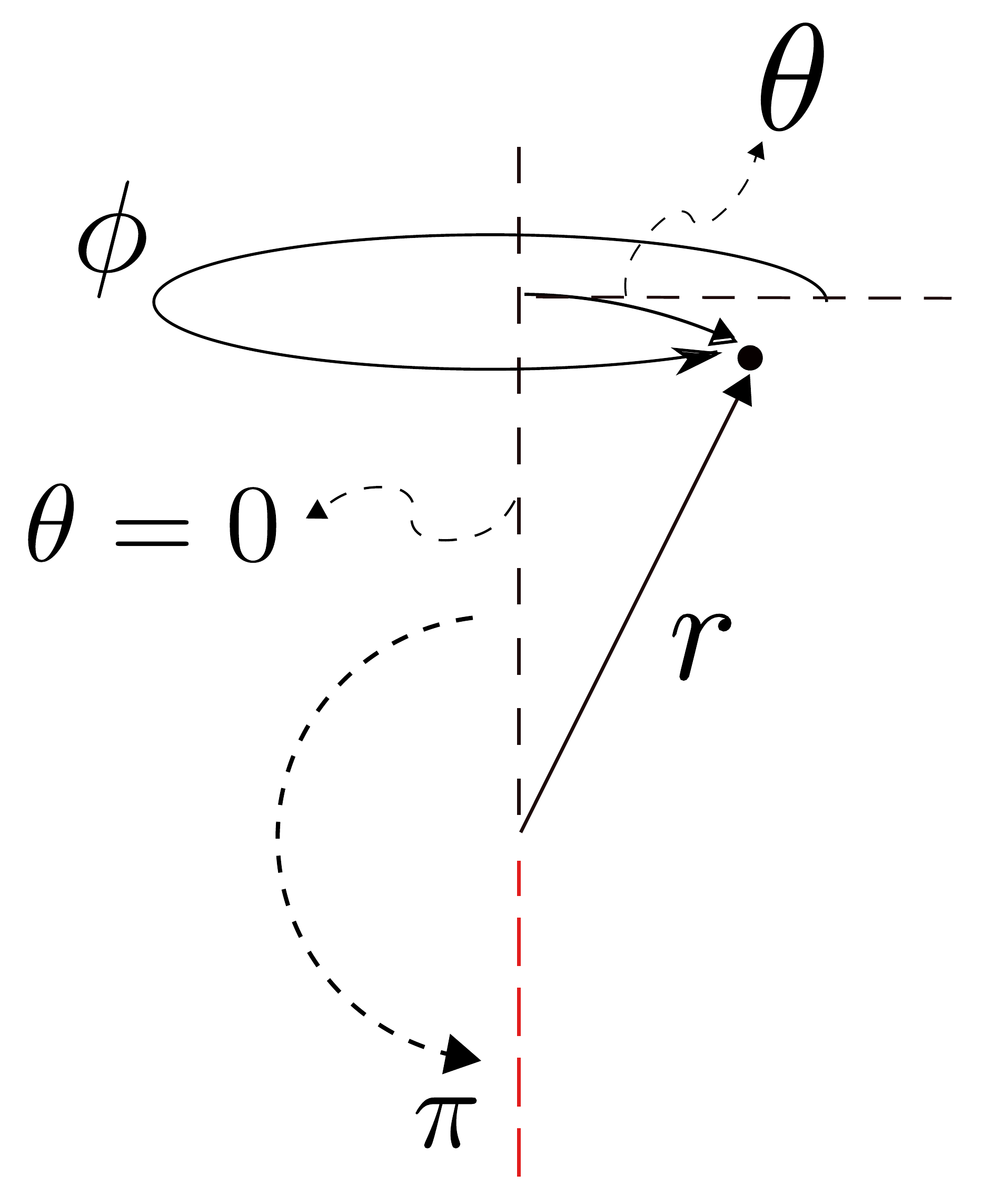} \label{cone}
\caption{\small The cone coordinates. {The limits of the angle $\theta$ are essential in defining the regular and outgoing functions.}}
\end{center}
\end{figure}
where $i_n$ is the modified spherical Bessel function.
By turning the summation into a contour integration with appropriate
poles on the real axis (see Fig.~\ref{ConeScalarCP}), we find
\begin{align} \label{PlnWavExpn}
  e^{-\kappa r \cos \Theta}&= \int_{0}^{\infty} d \lambda \ \lambda \tanh(\lambda \pi) k_{i \lambda -1/2}(\kappa r) P_{i\lambda -1/2}(\cos\Theta) \nonumber \\
  &=\sum_{m=-\infty}^{+\infty} \int_{0}^{\infty}
   d\lambda \ \lambda \tanh(\lambda
  \pi) (-1)^m  k_{i\lambda -1/2}(\kappa r) e^{im \phi} \times \cr
  & \null \hskip .9in \times P^{-m}_{i \lambda -1/2}(\cos \theta) e^{-im \psi} P^{m}_{i \lambda -1/2}(\cos a) \, ,
\end{align}
where $\theta$ and $\phi$ are the angles of $\bm{r}$ in spherical
coordinates, $a$ and $\psi$ are the corresponding angles for
$\bm{k}$, and $\Theta$ is the angle between $\bm{r}$ and $\bm{k}$.
In the last equation, we have used the identity \cite{Carslaw}
\begin{align} \label{LegendreAddThm2}
  P_{\nu}(\cos\Theta)
  =\sum_{m=-\infty}^{\infty}  (-1)^m e^{im(\phi-\phi')}
  P^{-m}_{\nu}(\cos \theta) P^{m}_{\nu}(\cos \theta')\, ,
  \hskip .2in (\theta+\theta'< \pi )
\end{align}
where $\cos\Theta= \cos \theta \cos \theta'+
\sin\theta\sin\theta'\cos(\phi-\phi')$.
Equation~(\ref{PlnWavExpn}) is only valid for Re$(\theta +
a) < \pi/2$, as can be seen from the asymptotic behavior
of Legendre functions of a large degree \cite{Carslaw, Abramowitz},
\begin{equation}\label{LegendreAsymptotics}
  P_{\nu}(\cos\theta) \,\sim\, \sqrt{\frac{2}{\nu \pi \sin \theta}}
  \sin\left((\nu+1/2)\theta +\pi/4\right)
  \hskip .5in \mbox{for large } |\nu|.
\end{equation}

By similar techniques, we can obtain the Green's function \cite{Carslaw,Felsen},
\begin{align} \label{GrnFnSph}
    G_0(&r, \theta, \phi;r' , \theta', \phi',i \kappa)=
    \frac{\kappa}{ 4 \pi}  \int_{0}^{\infty} d \lambda \ \lambda
    \tanh(\lambda \pi) k_{i \lambda -1/2}(\kappa r) k_{i\lambda
    -1/2}(\kappa r') P_{i \lambda -1/2}(-\cos\Theta),
\end{align}
which in its bilinear form becomes
\begin{align}
    G_0(r, \theta, \phi;r' , \theta', \phi', i\kappa)=
    &\frac{ \kappa }{4 \pi} \sum_{m=-\infty}^{+\infty}
    \int_{0}^{\infty} d\lambda \ \lambda \tanh(\lambda \pi) k_{i
    \lambda -1/2}(\kappa r) k_{i\lambda -1/2}(\kappa r') \times \nonumber \\
    & \times e^{i m(\phi -\phi')} P^{-m}_{i\lambda -1/2}(\cos
    \theta_<) P^{m}_{i\lambda -1/2}(-\cos \theta_>),
\end{align}
where $\theta_<$ ($\theta_>$) is the smaller (larger) of $\theta$ and
$\theta'$
and we used the identity of Eq.~(\ref{LegendreAddThm2}) in the form~\cite{Carslaw},
\begin{align} \label{LegendreAddThm1}
   P_{\nu}(-\cos\Theta) =   \sum_{m=-\infty}^{\infty}
   e^{im(\phi-\phi')} P^{-m}_{\nu}(\cos \theta) P^{m}_{\nu}(-\cos
   \theta')
   \hskip .5in \theta'>\theta \,.
\end{align}
The Green's function can be also obtained
by using the analog of Eq.~(\ref{OrthRln})~\cite{Carslaw},
\begin{equation}
  \frac{1}{ \pi} \int_{0}^{\infty} d \lambda \ \ \lambda \sinh(\lambda
  \pi) k_{i\lambda -1/2}(r) k_{i \lambda-1/2}(r')= \delta(r-r') \,.
\end{equation}
\begin{SCfigure}
\includegraphics[scale=0.37]{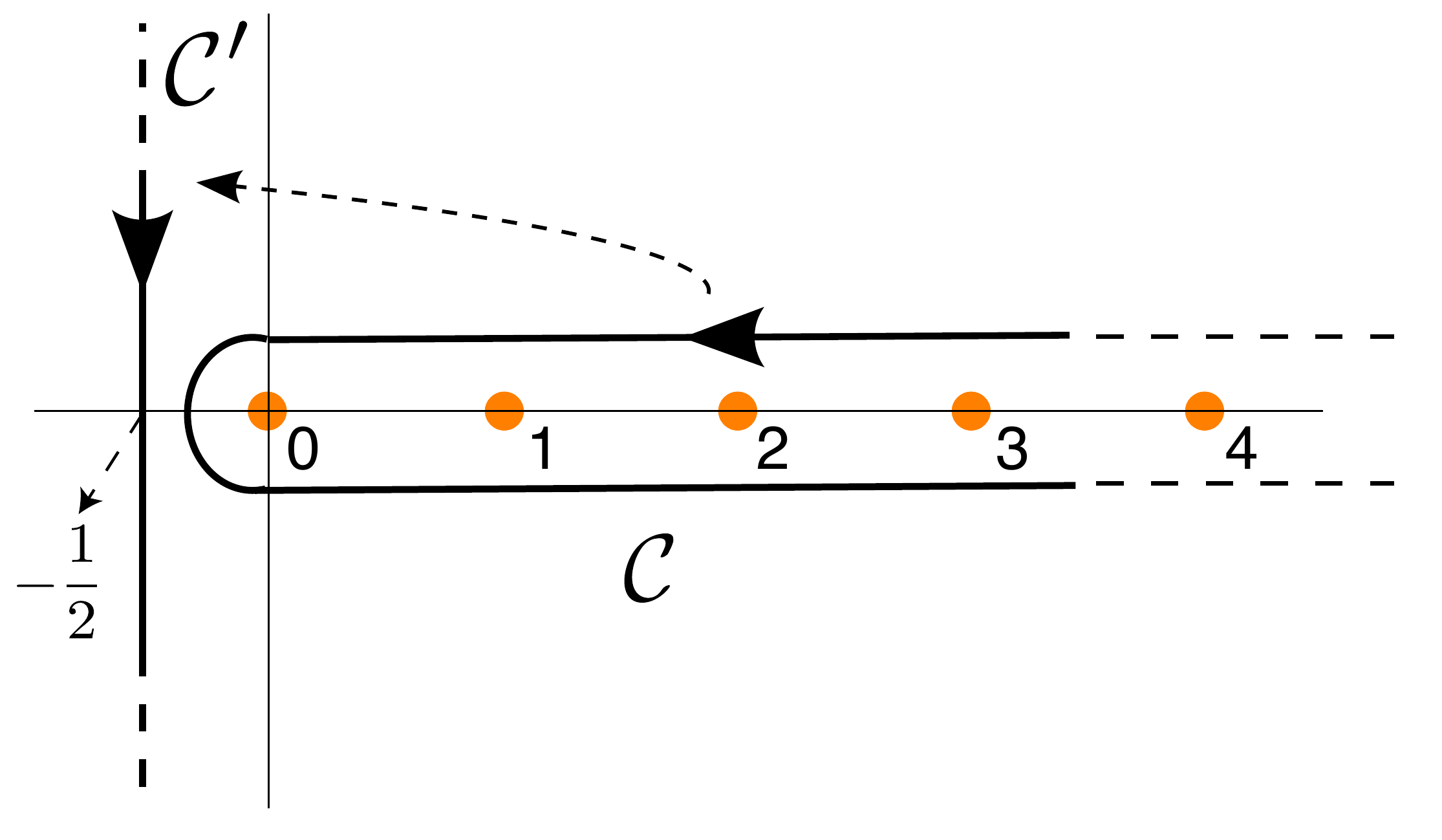}
\label{ConeScalarCP}
\caption{\small Analytic continuation to the imaginary axis {of the angular momentum} for {the cone in} the scalar case.}
\end{SCfigure}

From the Green's function, we identify the scattering
wavefunctions and the corresponding normalization factors,
\begin{align}
  &  \phi_{\kappa \lambda m}^{\rm reg}(r, \theta, \phi) = k_{i \lambda
  -1/2}(\kappa r) e^{i m\phi } P^{-m}_{i\lambda -1/2}(\cos \theta),
  \nonumber \\
  &  \phi_{\kappa \lambda m}^{\rm out}(r', \theta', \phi') = k_{i
\lambda -1/2}(\kappa r') e^{i m\phi' } P^{m}_{i\lambda -1/2}(-\cos
\theta'), \nonumber \\
  &  C_{\kappa \lambda m}= \frac{\kappa }{4\pi} \lambda
\tanh(\lambda \pi).
\end{align}
We note that $P_{i\lambda-1/2 }^{-m}(\cos \theta)$ is regular\footnote{ The superscript $-m$ of the regular wavefunctions is chosen to absorb a complicated normalization factor. The distinction between the regular and outgoing wavefunctions {arises from} the arguments of the Legendre functions, {not  their order}.}
everywhere except at $\theta=\pi$ (where it diverges)
while $P_{i\lambda-1/2 }^{m}(-\cos \theta)$ is regular at $\theta=\pi$
but not $\theta=0$, as we would expect for regular and outgoing
wavefunctions{,} respectively.

The $T$-matrices in the {above} basis can be easily found by matching
boundary conditions.  For a scalar field obeying Dirichlet and Neumann
boundary conditions {(labeled by {D and N respectively})} on {a cone of half opening-angle $\theta_{0}$}, we have
\begin{align}
  T_{{D} \, \lambda m}=-\frac{P^{-m}_{i\lambda -1/2}(\cos
  \theta_0)}{P^{m}_{i\lambda -1/2}(-\cos \theta_0)},  \nonumber \\
  T_{{N} \, \lambda m}= - \frac{{\frac{\partial}{\partial \theta_0}
  P^{-m}_{i\lambda -1/2}(\cos
  \theta_0)}}{{\frac{\partial}{\partial\theta_0} P^{m}_{i\lambda
  -1/2}(-\cos \theta_0)}}\,.
  \end{align}

Now we consider the   cone opposite to an infinite plate {obeying the same boundary conditions as the cone.} We initially
assume that the axis of the cone is perpendicular to the plate.
From Eq. (\ref{PlnWavExpn}), the conversion matrix takes the form
\begin{equation}
\label{eq:conversion_cone_plane}
  D_{\lambda m ,\bm{k}_{{\|}}}=\lambda \tanh(\lambda \pi) (-1)^m
  e^{-im \psi} P^{m}_{i \lambda -1/2}(\cos a) \, ,
\end{equation}
{where $\bm{k}_\|$ is the wave vector parallel to the plate.}
We note that Re$(a)=0$, so this expression is valid for $\theta <\pi/2$.

The $T$-matrix of the cone in the plane-wave basis is then
\begin{align}
  T_{\bm{k}_{{\|}}, \bm{k}'_{{\|}}}
   &=  \int_{0}^{\infty} d\lambda  \sum_{m=-\infty}^{+\infty}\int_{0}^{\infty}
   d\lambda' \int_{0}^{\infty} \sum_{m'=-\infty}^{+\infty}
   \frac{C_{\bm{k}_\|}}{C_{\kappa\lambda m}} D^\dagger_{\lambda m ,\bm{k}_\|
   } T_{\lambda m, \lambda'm'} D_{\lambda' m',\bm{k}'_\| \, .
   }
\end{align}
In the limit that $\theta \to \pi/2$, we can use the identity
\begin{equation}
  \int_{0}^{\infty} d \lambda \ \lambda \tanh(\lambda \pi) P_{i\lambda
  -1/2}^{m}(\cosh t) P_{i\lambda -1/2}^{-m}(\cosh u)=(-1)^m
  \delta(\cosh t -\cosh u) \, ,
\end{equation}
to show that the $T$-matrix reduces to $\mp (2 \pi)^2 \delta(\bm{k}_\|
-\bm{k}'_\|)$, where the upper (lower) sign corresponds to
Dirichlet (Neumann) boundary conditions.

Then{,} for the Dirichlet case we have
\begin{align} \label{Dir N-matrix}
  \mathcal{N}^{D}_{\lambda m,\lambda'm'}= -\frac{1}{2\pi} \lambda
  \tanh(\lambda \pi) \frac{\Gamma(i \lambda -m +1/2)}{\Gamma(i\lambda
  +m+1/2)} \frac{P^{m}_{i\lambda -1/2}(\cos \theta_0)}{P^{m}_{i\lambda
  -1/2}(-\cos \theta_0)} A^{{D}}_{\lambda m,\lambda'm'}(2\kappa d) \, {,}
\end{align}
where
\begin{align}
   A^{{D}}_{\lambda m,\lambda'm'}(x)
   &=\int_{0}^{2\pi}d\psi \int_{a=0}^{i\infty}d\cos a \
   e^{i(m-m') \psi} P_{i\lambda -1/2}^{-m}(\cos a)  P_{i\lambda'
   -1/2}^{-m'}(\cos a) e^{-x \cos a}\nonumber \\
   &=2\pi \delta_{m,m'} \int_{1}^{\infty} d \xi P_{i\lambda
   -1/2}^{m}(\xi)  P_{i\lambda' -1/2}^{m}(\xi) e^{-x \xi} \, .
\end{align}
We can then obtain the Casimir energy from
\begin{equation}
  \mathcal{E}_{D} = \frac{\hbar c}{2\pi} \int_{0}^{\infty} d\kappa\,
  \mbox{tr}\ln(\mathcal I_{\lambda m,\lambda'm'}- \mathcal{N}^{D}_{\lambda m,\lambda'm'})\,,
\end{equation}
where the trace is defined as tr\,$f_{ \lambda m ,
 \lambda'm'}=\int_{0}^{\infty} d\lambda \sum_{m=-\infty}^{\infty}
f_{ \lambda m,\lambda m}$.
 The Neumann $\mathcal{N}$-matrix can be readily found by making the
following substitution in Eq. (\ref{Dir N-matrix}),
\begin{equation}
\frac{P^{m}_{i \lambda-1/2}(\cos \theta_0)}{P^{m}_{i\lambda-1/2}(-\cos
\theta_0)} \hskip .1in \longrightarrow \hskip .1in
\frac{{\frac{\partial}{\partial \theta_0} P^{m}_{i\lambda -1/2}(\cos
\theta_0)}}{{\frac{\partial}{\partial\theta_0} P^{m}_{i\lambda
-1/2}(-\cos \theta_0)}} \quad .
\end{equation}

Next, we find an exact expression in the limit of small opening angle.
Since the $T$-matrix vanishes in this limit
for both the Dirichlet and Neumann case, we {expand the logarithm to first order} in the multiple reflections.
For the Dirichlet case, we only need to keep $m=0$ term in the limit of small angle,
since
\begin{equation}
 \frac{P_{\nu}(\cos \theta_0)}{P_{\nu}(-\cos
\theta_0)} {\,=\,} \frac{\pi}{2 \ln(\frac{\theta_0}{2})}
\frac{1}{\sin(\nu \pi)}+\mathcal O(\theta_0^2)\,,
\end{equation}
and the energy becomes
\begin{align} \label{Dirichlet Needle}
 \mathcal{E}{^{\rm cone}_{D}}{\,=\,} \frac{\hbar c}{2 \pi}& \frac{-1}{4
  \ln(\theta_0/2)}\int_{0}^{\infty} d\kappa \int_{1}^{\infty} d(\cos
  a) e^{-2 \kappa d \cos a}\int_{0}^{2\pi} d\psi
  \int_{0}^{\infty} d\lambda \ \lambda \frac{\tanh(\lambda
  \pi)}{\cosh(\lambda \pi)} \times \nonumber \\
  & \left\{ P_{i\lambda -1/2}(\cos a) P_{i\lambda -1/2}(\cos
  a )    \right\} +\mathcal O(\theta_0^2)+\cdots \nonumber \\
  &{\,=\,} -\frac{\hbar c}{16 \pi d} \frac{1}{|\ln(\theta_0 /2)|} +\mathcal O(\theta_0^2)+\cdots\, ,
\end{align}
{where the dots indicate corrections from higher reflections.}

For the Neumann case, in the limit of small angle,
\begin{equation}
\frac{{\frac{\partial}{\partial \theta_0}
P^{m}_{\nu}(\cos \theta_0)}}{{\frac{\partial}{\partial\theta_0}
P^{m}_{\nu}(-\cos \theta_0)}} {\,\sim\,}
\begin{cases}
\frac{(-1)^m \pi \Gamma(1 +\nu +m)}{\Gamma(m)\Gamma(m+1) \Gamma
(1+\nu-m) \sin(\nu \pi)}(\frac{\theta_0}{2})^{2m},
& \mbox{if }m \ne 0 \\
 -\frac{\pi \nu (\nu+1)}{\sin(\nu \pi )}(\frac{\theta_0}{2})^2, &
\mbox{if }m=0
\end{cases}
\end{equation}
so we need to consider $m=-1,0,1$. The Casimir Energy is
then given by
\begin{align}
 \mathcal{E}{^{\rm cone}_{N}}{\,=\,} \frac{\hbar c}{2 \pi}&\frac{\theta_0^2}{8}\int_{0}^{\infty}
  d\kappa \int_{1}^{\infty} d(\cos a) e^{-2 \kappa d \cos
  a}\int_{0}^{2\pi} d\psi \int_{0}^{\infty} d\lambda \ \lambda
  \frac{\tanh(\lambda \pi)}{\cosh(\lambda \pi)} \times \nonumber \\
  & \left\{ -P_{i\lambda -1/2}(\cos a) P_{i\lambda -1/2}(\cos
  a ) (\lambda^2+1/4) +2 P^1_{i\lambda -1/2}(\cos a)
  P^1_{i\lambda -1/2}(\cos a)   \right\}+\mathcal O(\theta_0^4)+\cdots\nonumber \\
  &{\,=\,} -\frac{\hbar c}{24\pi d} \theta_0^2 +\mathcal O(\theta_0^4)+\cdots,
\end{align}
where in the last line we have used integral identities {given in Ref.~}\cite{Felsen}{. Again, the dots represent higher reflections}.

\subsection{Electromagnetic field}
\label{subsec:electriccone}
Because the cone does not have translation symmetry, the results for
electromagnetism can no longer be obtained as a simple combination of
two scalar problems.  Using the same techniques {as in the
  scalar case}, however, we can
obtain the electromagnetic Green's function for the cone {in a useful
  form} by analytic continuation of the {usual representation in terms
  of spherical partial waves.} As far as we know, this Green's
function has not been obtained previously in the literature, so we
{derive} it in detail in Section~\ref{AppendixGreenFn}.  We find that
this case contains an additional subtlety, connected to the absence of
the $l=0$ mode in electromagnetism (see Fig.~\ref{ConeElecCP}).  The
Green's function takes the form
\begin{align}\label{El Grn Fn}
  \mathbb{G}_0
  =& - \frac{\kappa}{4 \pi}\sum_{m=-\infty}^{\infty}
  \int_{0}^{\infty} d\lambda  \ \lambda \tanh(\lambda
  \pi)\frac{1}{\lambda^2+1/4}(\mathbf{M}_{i \lambda-1/2, m }^{\rm out}
  \mathbf{M}_{i \lambda-1/2, m }^{{\rm reg}*} -\mathbf{N}_{i \lambda-1/2, m
  }^{\rm out}\mathbf{N}_{i \lambda-1/2, m }^{{\rm reg}*}) \nonumber \\
  &- \frac{\kappa}{4 \pi}\sum_{m=-\infty, m \ne 0}^{\infty}
  \Gamma(|m|)\Gamma(|m|+1) \mathbf{R}_{0, m}^{\rm out}\
  \mathbf{R}_{0,m}^{{\rm reg}*}\, ,
\end{align}
where we have defined the outgoing and regular wave{ function}s
\begin{align}\label{El Fns}
  & \mathbf{M}_{i \lambda-1/2, m }^{\rm out}= {\boldsymbol \nabla} \times
  k_{i\lambda-1/2}(\kappa r) P_{i \lambda -1/2}^{m}(-\cos\theta) e^{i m
  \phi} \mathbf{r}\, , \nonumber \\
  &\mathbf{M}_{i \lambda-1/2, m }^{\rm reg}= {\boldsymbol \nabla} \times
  k_{i\lambda-1/2}(\kappa r) P_{i \lambda -1/2}^{-m}(\cos\theta) e^{i m
  \phi} \mathbf{r}\, , \nonumber \\
  & \mathbf{N}_{i \lambda-1/2, m }^{\rm out}=\frac{1}{\kappa} {\boldsymbol \nabla}
  \times {\boldsymbol \nabla} \times k_{i\lambda-1/2}(\kappa r) P_{i \lambda
  -1/2}^{m}(-\cos\theta) e^{i m \phi} \mathbf{r}\, , \nonumber \\
  &\mathbf{N}_{i \lambda-1/2, m }^{\rm reg}= \frac{1}{\kappa} {\boldsymbol \nabla}
  \times {\boldsymbol \nabla} \times k_{i\lambda-1/2}(\kappa r) P_{i \lambda
  -1/2}^{-m}(\cos\theta) e^{i m \phi} \mathbf{r}\, , \nonumber \\
  &\mathbf{R}_{0, m}^{\rm out}= k_{0}(\kappa r) \ \mathbf{r} \times {\boldsymbol \nabla}
  P_{0}^{ - |m|}(- \cos\theta)e^{im \phi}\, ,\nonumber \\
  &\mathbf{R}_{0,m}^{\rm reg}=k_{0}(\kappa r) \ \mathbf{r} \times {\boldsymbol \nabla}
  P_{0}^{ - |m|}( \cos\theta)e^{im \phi}\quad .
\end{align}
In this decomposition we have obtained the usual magnetic (transverse
electric) modes ${M}$ and electric (transverse magnetic) modes
${N}$, but we also have an additional set of {discrete} modes
${R}$, arising from the pole at the origin in the contour
integral.

From the Green's function, we can also read off the normalization
coefficients
\begin{align}
  & C^{M}_{\lambda m}=-C^{N}_{\lambda m }= - \frac{\kappa}{4 \pi}
  \frac{1}{\lambda^2 +1/4}\lambda \tanh\lambda \pi, \nonumber \\
  & C^{R}_{m}= - \frac{\kappa}{4 \pi} \Gamma(|m|)\Gamma(|m|+1).
\end{align}
For a perfect {reflector}, the $T$-matrix is diagonal in the space of
the {${M}$, ${N}$, and ${R}$ modes,}{\footnote{{In Fig.~1, $M$, $N$ and $R$ modes are referred to as magnetic ($M$), electric ($E$) and the ghost ($Gh$) fields respectively.}}}
\begin{align}
{T_{M\,\lambda m}}&=-\frac{\partial_{\theta_0} P_{i \lambda -1/2}^{
-m}(\cos \theta_0)}{\partial_{\theta_0} P_{i \lambda -1/2}^{m}(-\cos
\theta_0)}, {\hskip .3in -\infty<m<\infty} \nonumber \\
{ T_{N\,\lambda m}}&=-\frac{ P_{i \lambda -1/2}^{ -m}(\cos \theta_0)}{
P_{i \lambda -1/2}^{ m}(-\cos \theta_0)},  {\hskip .3in -\infty<m<\infty}\nonumber \\
{T_{R\, m}}&=\frac{ P_{0}^{ -|m|}(\cos \theta_0)}{P_{0}^{ -|m|}(-\cos
\theta_0)} = \left(\tan\frac{\theta_0}{2}\right)^{2 |m|}, \hskip .3in
m\ne 0.
\end{align}
\begin{SCfigure}
 \includegraphics[scale=.3]{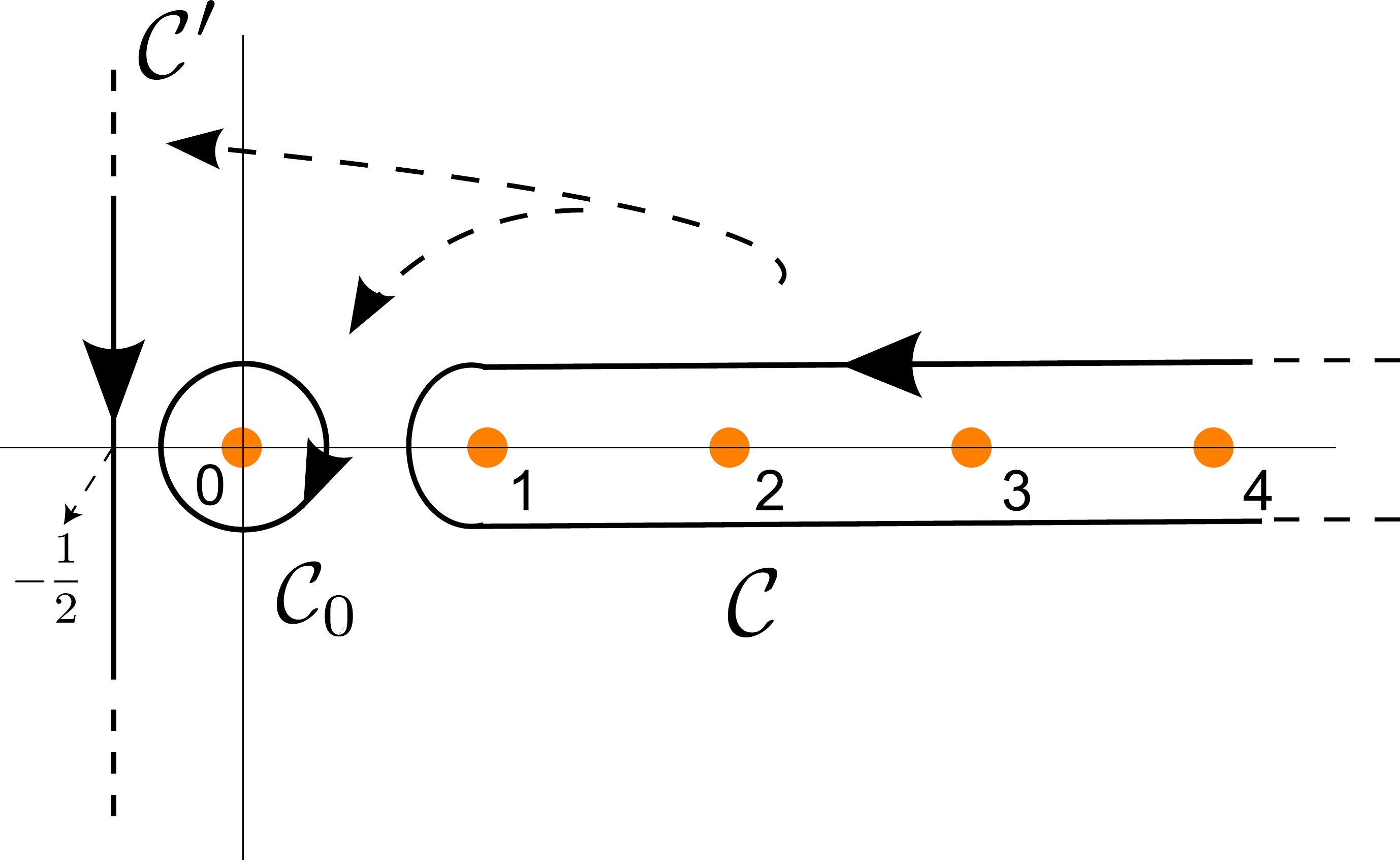}
\label{ConeElecCP}
\caption{\small Analytic continuation to imaginary axis {of the angular momentum} {for the EM case}.}
\end{SCfigure}

We will also need the conversion matrices between the {cone}
and plane wave bases,
\begin{align}
  \mathbf{M}^{\rm reg}_{\bm{k}_{\|}}=
  & \sum_{m=-\infty}^{\infty}\int_{0}^{\infty}d \lambda \ D^{M
  \bm{k}_{\|}}_{M\lambda m} \mathbf{M}_{i \lambda-1/2, m
  }^{\rm reg}+D^{M \bm{k}_{\|}}_{N\lambda m} \mathbf{N}_{i
  \lambda-1/2, m }^{\rm reg}+ \sum_{m=-\infty, m\ne 0}^{\infty} D^{M
  \bm{k}_{\|}}_{R\, m} \mathbf{R}_{0, m }^{\rm reg}, \nonumber \\
  \mathbf{N}^{\rm reg}_{\bm{k}_{\|}}=&
  \sum_{m=-\infty}^{\infty}\int_{0}^{\infty} d\lambda \ D^{N
  \bm{k}_{\|}}_{M \lambda m} \mathbf{M}_{i \lambda-1/2, m
  }^{\rm reg}+D^{N \bm{k}_{\|}}_{N\lambda m} \mathbf{N}_{i
  \lambda-1/2, m }^{\rm reg}+ \sum_{m=-\infty, m\ne 0}^{\infty} D^{N
  \bm{k}_{\|}}_{R\, m} \mathbf{R}_{0, m }^{\rm reg}\,,
\end{align}
which can be obtained by analytic continuation of the expansion of a
plane wave in vector spherical harmonics, yielding
\begin{align} \label{ConvMatrixElecCone}
  &D^{M \bm{k}_{\|}}_{M\lambda m}= -\frac{\lambda \tanh(\lambda
  \pi) }{\lambda^2 +1/4}\sinh \alpha \, e^{-im \psi} P
    _{i \lambda
  -1/2}^{'m}(\cosh \alpha), \nonumber \\
  &D^{M \bm{k}_{\|}}_{N\lambda m} =-\frac{\lambda \tanh(\lambda
  \pi) }{\lambda^2 +1/4} i m \frac{1}{\sinh \alpha} e^{-i m \psi} P_{i
  \lambda -1/2}^{m}(\cosh\alpha),\nonumber \\
  &D^{M \bm{k}_{\|}}_{R\, m} = -|m| \Gamma(|m|)
  \Gamma(|m|+1)\frac{1}{\sinh \alpha}e^{-i m \psi} P_{0}^{ - |m|}(\cosh
  \alpha) ,\nonumber \\
  &D^{N \bm{k}_{\|}}_{M \lambda m}= -D^{M
  \bm{k}_{\|}}_{N\lambda m}, \nonumber \\
  &D^{N \bm{k}_{\|}}_{N\lambda m}=D^{M
  \bm{k}_{\|}}_{M \lambda m},\nonumber \\
  &D^{N \bm{k}_{\|}}_{R\, m}=  i m \Gamma(|m|)
  \Gamma(|m|+1)\frac{1}{\sinh \alpha}e^{-i m \psi} P_{0}^{ - |m|}(\cosh
  \alpha)
\end{align}
where $a=i\alpha$ and $\psi=\angle
\mathbf k_\|$ are the angles of the plane wave in spherical
coordinates, {\it i.e.}, $\sinh \alpha= |\mathbf k_{\|}|/ \kappa$.

We can now use these results to {recast} the $T$-matrix for the cone in
the plane-wave basis appropriate for scattering from a
plate.  Symbolically and schematically, this transformation can be written as
\begin{equation}
T_{P, P'}= \sum_{m} \int_{0}^{\infty} d \lambda \ \sum_{Q} C_P
{D^{\dagger}}^{ P}_{Q}\, \frac{1}{C_Q}T_Q D^{P'}_{Q}\,,
\end{equation}
where $P,P'\in \{M,N\}$ are the wave functions in the plane wave basis
whereas $Q\in\{M, N,
R\}$ are defined in the conical basis.
The $\mathcal{N}$-matrix is then
\begin{equation}
\mathcal{N}_{P\bm{k}_{\|}, P'\bm{k}'_{\|}}=\frac{1}{(2 \pi)^2
} e^{-2 \kappa d \cosh{\alpha}} \, r^{P} \ T_{P\bm{k}_{\|},
P'\bm{k}'_{\|}},
\end{equation}
where
\[
r^P=
\begin{cases}
  -1 \hskip .2in &\mbox{if    } P=M\ (\rm{Magnetic\  mode}) \\
  1 \hskip .2in &\mbox{if    } P=N  \ (\rm{Electric\  mode}).
\end{cases}
\]

We first consider a cone of small opening angle. In this
limit, it is sufficient to consider $m=0$ of the mode ${N}$, in which case only one component of the conversion matrix,
$D^{N}_{N}$, is nonzero.
The $T$-matrix then simplifies to
\begin{align}
  T_{N \bm{k}_{\|}, N\bm{k}'_{\|}}=
  \frac{ \pi^2}{ |\ln{\frac{\theta_0}{2}}|} \int_{0}^{\infty} d \lambda
  \ \frac{\lambda \tanh{\lambda
  \pi}}{\lambda^2+1/4}\frac{1}{\cosh{\lambda \pi}} \sinh \alpha \
  P_{i \lambda-1/2}^{'}(\cosh \alpha) \sinh\alpha' \ P_{i
  \lambda-1/2}^{'}(\cosh  \alpha')+ \mathcal O(\theta_0^2).
\end{align}
To leading order, it suffices {to consider only the first term
  of the multiple scattering}
expansion. {Then to leading order as $\theta_{0}\to 0$ the electromagnetic Casimir energy of a cone is given by}
\begin{align} \label{electromagnetic Needle}
  \mathcal{E}^{{\rm cone}}_{{\rm EM}}&{\,=\,} - \frac{\hbar c}{2 \pi} \int_{0}^{\infty} d\kappa
  \int_{1}^{\infty} d\cosh\alpha \int_{0}^{2 \pi} d\psi \ \frac{1}{(2
  \pi)^2} e^{-2 \kappa d \cosh\alpha}T_{N\bm{k}_{\|}, N
  \bm{k}_{\|}} +\cdots \nonumber \\
  &{\,=\,} -\frac{\hbar c}{2 \pi} \frac{1}{8d}
  \frac{1}{|\ln{\frac{\theta_0}{2}}|} \int_{1}^{\infty}
  \frac{d\cosh\alpha}{\cosh{\alpha}} \int_{0}^{\infty} d\lambda \ \lambda
  \frac{\tanh{\lambda \pi} }{\cosh{\lambda \pi }}
  \frac{1}{\lambda^2+1/4}P^1_{i \lambda-1/2}(\cosh\alpha) P^1_{i
  \lambda-1/2}(\cosh\alpha) +\mathcal O(\theta_0^2) +\cdots\nonumber \\
  &{\,=\,} -\frac{\hbar c }{d} \ \frac{\ln 4-1}{16\pi}
  \frac{1}{|\ln{\frac{\theta_0}{2}}|}+\mathcal O(\theta_0^2) +\cdots,
\end{align}
where the dots represent corrections from higher reflections. This result {has the same dependence
  on $d$ and $\theta_0$ as the one for} the Dirichlet cone at
small angle $\theta_0$ {[see Eq.~(\ref{Dirichlet Needle})]}, though the electromagnetic
result is smaller by about 40$\%$.

For arbitrary opening angle, we compute the Casimir energy by keeping the first two terms in the multiple reflection expansion. The results are accurate within one percent and are not modified by higher order terms at the level of accuracy {shown in Fig.~3}.

\subsection{Tilted cone}
\label{subsec:tiltedcone}
It is straightforward to allow the cone to tilt. Here we {present}
the Casimir energy of a tilted needle { --- a cone in the limit of
  vanishing opening angle --- } opposite a plate.  Since the
$T$-matrix vanishes for small opening angle, we again keep only the
first term in the multiple reflection expansion.
To introduce a tilt,
\begin{figure}[h]
  \begin{center}
  \includegraphics[scale=.2]{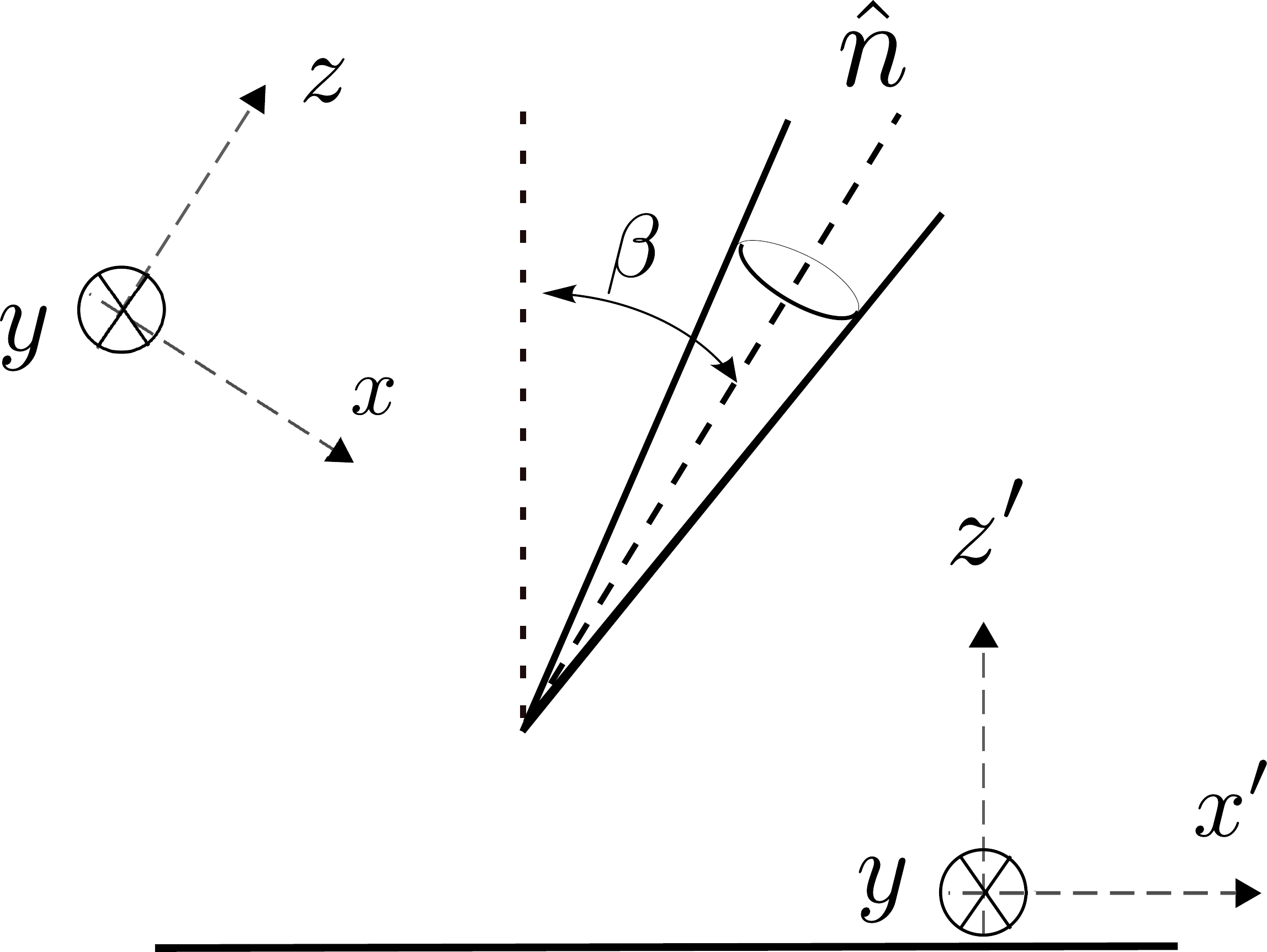}
\caption{\label{Tilted Needle}\small {The configuration of} a tilted needle {opposite an infinite} plane.}
\end{center}
\end{figure}
{it suffices to} consider the plane wave in a rotated basis.  For the
Dirichlet case, the conversion matrix {is given by Eq.~\eqref{eq:conversion_cone_plane},}
where $a$ and $\psi$ give the spherical polar and azimuthal angles of $\bm{k}$,
respectively, with respect to $\hat{\bm{n}}$, the symmetry axis of the cone.
These are given in terms of spherical coordinates by
\begin{align} \label{rotation}
  \cos a&=\cos a'\cos\beta+ \sin  a'\cos\psi'\sin \beta, \cr
 \tan\psi&=\frac{\tan \psi'}{\cos \beta -\cot  a'\csc\psi'\sin \beta}
\end{align}
where $\beta$ is the tilt of the cone, and $a'$ and $\psi'$ are the
angular coordinates of $\bm{k}$ with respect to the $x', y', z'$ axes,
which form a convenient basis for the plate{, see
  Fig.~\ref{Tilted Needle}.} These angles are defined according to
\begin{align}
\cos a'=  \frac{\sqrt{\kappa^2+\bm{k}_\|^2}}{\kappa}, \cr
\psi'=\angle \bm{k}_\|.
\end{align}
Since $a'$ is imaginary, we define $a' = i \alpha'$, so that
\begin{align}
\cos  {a}& =
\cosh\alpha'\cos\beta+ i\sinh \alpha'\sin\psi' \sin \beta \cr
\tan\psi &=
\frac{\tan \psi'}{\cos \beta +i\coth \alpha'\csc\psi'\sin \beta}.
\end{align}
{For the Dirichlet cone in the limit of vanishing opening angle --- the Dirichlet needle --- tilted at an angle $\beta$ from the normal to the plate, the Casimir energy becomes,}
\begin{align}
  \mathcal{E}{_{ D}^{\rm needle}}&{\,=\,} \frac{\hbar c}{2 \pi} \frac{-1}{4
  \ln(\theta_0/2)}\int_{0}^{\infty} d\kappa \int_{1}^{\infty} d(\cosh
  \alpha') e^{-2 \kappa d \cosh \alpha'}\int_{0}^{2\pi} d\psi'
  \int_{0}^{\infty} d\lambda \ \lambda \frac{\tanh(\lambda
  \pi)}{\cosh(\lambda \pi)} \times \nonumber \\
  & {\times\,\,  P_{i\lambda -1/2}(\cos  {a}) P_{i\lambda
  -1/2}(\cos{a}^*)} +\mathcal O(\theta_0^2)+\cdots\,,    \nonumber \\
    &{\,=\,}-\frac{\hbar c}{16\pi d} \frac{1}{|\ln(\theta_0 /2)|}
  \frac{1}{\cos \beta} +\mathcal O(\theta_0^2)+\cdots.
\end{align}
Again the dots represent corrections from higher reflections.

For electromagnetism, similar changes should be made in the conversion
matrix {to obtain} the energy for a tilted needle,
\begin{align}
\mathcal{E}{_{\rm EM}^{\rm needle}}&{\,=\,} \frac{\hbar c}{2 \pi} \frac{-1}{4 \ln(\theta_0/2)}
\int_{1}^{\infty}d\kappa \int_{1}^{\infty} d(\cosh\alpha') e^{-2
\kappa d \cosh \alpha'}\int_{0}^{2\pi} d\psi' \int_{0}^{\infty}
d\lambda \ \lambda \frac{\tanh{\lambda \pi} }{\cosh{\lambda \pi }}
\frac{1}{\lambda^2+1/4}\times
\nonumber \\
&{\times\,\, P^1_{i \lambda-1/2}(\cos{a}) P^1_{i
\lambda-1/2}(\cos{a}^*) } +\mathcal O(\theta_0^2)+\cdots \nonumber \\
&{\,=\,} -\frac{\hbar c }{16 \pi d}  \frac{1}{|\ln(\theta_0 /2)|}
g(\beta) \frac{1}{\cos \beta} +\mathcal O(\theta_0^2)+\cdots \, ,
\end{align}
where $g(\beta)$ is given by
\begin{align}
 g(\beta)=& \int_{0}^{2\pi} \frac{d \psi'}{2 \pi} \int_{1}^{\infty}
 d\cosh \alpha' \frac{1}{\cosh^2\alpha'
 }\left|\frac{1-\cos {a}}{1+\cos {a}}\right| \cr
 =&\int_{0}^{2\pi} \frac{d \psi'}{2 \pi} \int_{0}^{\infty} d \alpha'
 \frac{\tanh\alpha'}{\cosh\alpha' }\sqrt{\frac{(\cosh \alpha'
 \cos\beta -1)^2+(\sinh \alpha'\sin \psi'\sin\beta )^2 }{(\cosh
 \alpha' \cos\beta +1)^2+(\sinh \alpha'\sin \psi'\sin\beta )^2}}.
\end{align}
The function $g(\beta)$ goes monotonically from $g(0)=\ln 4 -1$ to $g(\pi/2)=1$
{(see Fig.~{3})}.

\section{Finite temperature}
\label{sec:thermal}

{At finite temperatures the integral over
  the imaginary wave number $\kappa$ is replaced by a sum over
  Matsubara wavenumbers, {$\kappa_{n}= \frac{2 \pi k_B T}{\hbar c }
    n$}.  To gauge the importance of finite temperature effects, we
  report studies of two cases: 1) A {perfectly} conducting knife edge (a wedge in
  the limit of zero opening angle) opposite a conducting plate; and 2)
  A {perfectly} conducting needle (a cone in the limit of zero opening angle)
  either aligned vertically above a conducting plate or inclined at an
  angle $\beta$ from the vertical.}

\subsection{Knife edge}
\label{subsec:thermalknife}
We consider a knife edge
with an arbitrary tilt $\phi_0$. It would be
straightforward to generalize the following results to nonzero opening
angle.
In Section~\ref{sec:wedge}, we found the Casimir energy of this system by reducing the problem to a two dimensional one. To include the effects of finite temperature,
the original frequency $ \kappa$ must be restored and summed
over Matsubara frequencies. We consider a single reflection (back and
forth) between the two objects. As described in Section~\ref{sec:wedge}, this term gives a good
approximation to the {exact} result.  The free energy is given by
\begin{equation}
  {\mathcal{F}}= -k_B T \frac{L_z}{2 \pi} {\sum_{
  \kappa_n \ge 0}}' \ \mbox{tr}\mathcal{N}+\cdots \,,
\end{equation}
where $T$ is the temperature and the primed sum
indicates that the $n=0$ term is counted with a weight of $1/2$. The dots represent higher terms in the multiple reflection expansion.
For scalars subject to Dirichlet and Neumann boundary conditions,
tr\,$\mathcal{N}$ becomes
\begin{equation}
\label{trNthermal}
  \mbox{tr}\,\mathcal N_{D/N}^{ \substack{ \rm knife\\  \rm edge}}= \int dk_x dk_z \ \frac{1}{4 \pi} e^{-2 d \sqrt{\kappa_n^2+k_x^2+k_z^2}}\frac{1}{\sqrt{\kappa_n^2+k_x^2+k_z^2}}\left(\pm \frac{\sqrt{\kappa_n^2+k_z^2}}{\sqrt{\kappa_n^2+k_x^2+k_z^2}}+\frac{1}{\cos \phi_0}\right),
\end{equation}
so that for electromagnetism the first term {cancels between Dirichlet and Neumann modes} and we get an
overall factor of 2 (summing over both modes). {This} yields {the EM Casimir free energy}
{to first order in the multiple scattering expansion,}
\begin{align}\label{wedge-Finite T2}
  {\mathcal{F}}_{\rm EM}^{ \substack{ \rm knife\\  \rm edge}} =-\frac{\hbar c L_z}{16 \pi^2}\frac{1}{\cos \phi_0}
  \frac{1}{d^2} { \frac{d}{\lambda_T} \coth  \frac{d}{\lambda_T} }+\cdots\,.
\end{align}
where {$\lambda_T=\frac {\hbar c }{2 \pi k_B T}$}.  Expanding the {\it force} at low temperature, we find
\begin{equation}
F_{\rm EM}^{ \substack{ \rm knife\\  \rm edge}} =-\frac{\hbar c L_z}{8 \pi^2}  \frac{1}{\cos \phi_0} {\frac{1}{d^3} \left[
1 + \frac{1}{45} \left(\frac{d}{\lambda_T}\right)^4+ \mathcal O\left(\left(\frac{d}{\lambda_T}\right)^6\right)\right]+\cdots\,.}
\end{equation}
Here we have found the contribution from the first reflection which shows that the force {depends only weakly {on $T$} at low temperature.} The dots represent the contribution from higher reflections, but as argued before, the latter is significantly smaller.}
It is interesting to note that the last equations are quite similar to the {result from} two parallel plates. To the first order in multiple reflections, the free energy of two parallel {perfectly reflecting} plates at temperature $T$ is
\begin{align}
  {\mathcal{F}}_{\rm EM}^{ \substack{ \rm parallel\\  \rm plates}}&=k_B T \frac{L_x L_z}{8\pi}
  \frac{\partial}{\partial d} \left(\frac{1}{d} \coth {\frac{d}{\lambda_T}} \right)+\cdots.
\end{align}
When $\epsilon=\pi/2-\phi{_0}$ is small, the result for the wedge agrees with the result for parallel plates, if $\epsilon$ is identified correctly in terms of temperature and other parameters of the problem.

{In contrast, {our numerical results follwing from Eq.~(\ref{trNthermal}) indicate that} thermal corrections for a scalar field obeying Dirichlet or Neumann boundary conditions on a knife edge are proportional to temperature {with a} power close to 3. This points to the {complex} interplay of temperature dependence and geometry,
which was previously noted in Ref.~\cite{Klingmueller08}.

\subsection{Needle}
\label{subsec:thermalneedle}
While temperature corrections can be computed for cones of arbitrary
opening angle, we can obtain a closed-form
analytical formula  for a needle. Hence we focus on the latter case below.  Our calculation
exploits the fact that the $T$-matrix tends to zero for a sharp needle
and thus the leading term in the multiple reflection expansion is
sufficient.  We will see that at any nonzero temperature, the free
energy diverges in the infrared.  This divergence doesn't
depend on the objects' separation, however, so the force remains
finite.

{To} first
order in multiple reflections, the free energy {is}
\begin{align}
  {\mathcal{F}}&=k_B T {\sum_{\kappa_n \ge 0}}' \
  \mbox{tr\,}\mathcal{N} +\cdots\,.
\end{align}
For Dirichlet {boundary condition}, this gives
\begin{align}
  {\mathcal{F}}^{\rm needle}_{D}&=-\frac{k_B T}{4|\ln \theta_0/2| }
  {\sum_{n=0}^{\infty}}'
  \int_{1}^{\infty} dx \ \frac{1}{x} \exp( -2\kappa_n d \, x) \,+\mathcal O(\theta_0^2) +\cdots\,.
\end{align}
From the last equation, it is clear that the
energy becomes infinite for any finite $T$ due to the $n=0$ term
of the sum.  However, if we take the derivative with respect to $d$ to
get the force, we obtain
\begin{align}
  F^{\rm needle}_{ D}&=-\frac{k_B T}{4|\ln \theta_0/2| }
  {\sum_{n=0}^{\infty}}' 2 \kappa_n \int_{1}^{\infty} dx \  \exp(
  -2\kappa_n d \,x)+\mathcal O(\theta_0^2)+\cdots\,.
\end{align}
Note that
the integral must be computed
first; only then can the sum be performed.  If we separate
the summation into the first term ($n=0$) and
a sum over all $n\ge 1$, however, the latter sum commutes with the
integral. We then obtain
\begin{align}
  F^{\rm needle }_{ D}
           =-\frac{k_B T}{8|\ln \theta_0/2| d} \coth {\frac{d}{\lambda_T}} +\mathcal O(\theta_0^2)+\cdots \, .
\end{align}
This {is equivalent} to Eq. (\ref{Dirichlet Needle}) for $T\to 0$, while
it becomes linear in $T$ for $T\to \infty$.  An interesting feature is the
leading thermal correction for small temperatures. By expanding the last
expression, we find
\begin{equation}
  F^{\rm needle}_{D}=-\frac{\hbar c }{16 \pi |\ln \theta_0/2| }
  {\frac{1}{d^2} \left[ 1+\frac{1}{3} \left(\frac{d}{\lambda_T}\right)^2+\mathcal O \left(\left(\frac{d}{\lambda_T}\right)^4\right) \right]} +\mathcal O(\theta_0^2)+\cdots  \,.
\end{equation}
{This shows that} thermal corrections to the force {have} a significantly larger effect than {in all} other known examples, most notably two parallel plates\footnote{Relatively large thermal corrections, $\mathcal O\left(T^3\right)$, have been reported for Dirichlet half-plate geometry in Ref.~\cite{Klingmueller08}.}.

For electromagnetism, we find the result
\begin{align}
F^{\rm needle}_{\rm EM}= -\frac{k_B T}{8|\ln \theta_0/2| d} \left(1+ {\frac{d}{\lambda_T}}
\int_{1}^{\infty} dx \  \frac{x-1}{x+1} \frac{1}{ \sinh{^{2}} ({d/\lambda_T}{)} \, x}\right)+\mathcal O(\theta_0^2)+\cdots  .
\end{align}
Again this {is equivalent} to Eq. (\ref{electromagnetic Needle}) for
$T\to 0$ and it {scales} linearly with $T$ at large temperature. Expanding this equation at low temperature, we obtain the leading thermal correction,
\begin{equation}
  F_{\rm EM}^{\rm needle}= -\frac{\hbar c }{16 \pi}
  \frac{1}{|\ln{\frac{\theta_0}{2}}|} \left(\frac{\ln 4-1}{d^2}-\frac{2}{3{\lambda_T^2}}  \ln(2  d/{\lambda_T})  {+0.810\,} {\frac{1}{\lambda_T^2}}+\cdots\right)+\mathcal O(\theta_0^2)+\cdots   \, ,
  \label{thermalEM}
\end{equation}
{where the coefficient of the third term equals $\frac{1}{3} (-1 - 2 \gamma -  \ln 4 + 24 \ln A)$, in which $\gamma$ is Euler's constant and $A$ the Glaisher-Kinkelin constant.}
Interestingly, the first thermal correction {is {again} proportional to} a (second) power of $T$ {but now also} multiplied by the logarithm of  {$d/\lambda_T$.}

{We can extend these results for a needle tilted by an angle $\beta$. }The Dirichlet free energy is just multiplied by $1/\cos \beta$,
\begin{align}
  F_{ D}^{\substack{\rm tilted\\ \rm needle}}=-\frac{k_B T}{8|\ln \theta_0/2|d}\coth( {d/\lambda_T}) \frac{1}{\cos
  \beta}+\mathcal O(\theta_0^2)+\cdots\,.
\end{align}
For electromagnetism, the $\beta$-dependence changes to
\begin{align}
  F_{\rm EM}^{\substack{\rm tilted\\ \rm needle}}=-\frac{\hbar c }{16 \pi |\ln \theta_0/2|d^2 } \frac{g(\beta, {d/\lambda_T})}{\cos
  \beta}+\mathcal O(\theta_0^2)+\cdots\,,
\end{align}
where $g(\beta, {d/\lambda_T})$ is defined as
\begin{equation}
\label{thermalEMtilt}
  g(\beta,  {d/\lambda_T})= {\frac{ d}{\lambda_T} + \left(\frac{d}{\lambda_T}\right)^2 } \int_{0}^{2\pi}
  \frac{d\psi'}{2\pi}\int_{1}^{\infty} dx  \ \sqrt{\frac{(x\cos
  \beta-1)^2+ (x^2-1)\sin^2 \beta \sin^2 \psi'}{(x\cos
  \beta+1)^2+ (x^2-1)\sin^2 \beta \sin^2 \psi'}}\
  \frac{1}{\sinh^2 ( {d/\lambda_T}) x}.
\end{equation}
{From this equation, the thermal correction at low temperature {changes} from $\mathcal{O}(T^2 \ln T)$ for a vertical (EM) needle to $\mathcal{O}(T^2)$ when the needle becomes almost parallel to the plane.}
{Eq.~(\ref{thermalEM}) and (\ref{thermalEMtilt}) predict  large temperature corrections to the Casimir force at room temperature, as shown in Fig.~{3.}

  The striking dependence on geometry of both the analytic form
  and the {size} of finite temperature corrections was suggested in
  Ref.~\cite{Scardicchio:2005di} and studied for a scalar field
  obeying Dirichlet boundary conditions in Ref.~\cite{Klingmueller08}.   Here we found that thermal corrections are relatively
  small for a wedge/knife-edge (along with parallel plates) while they
  are significantly larger for a needle.}
\\

\section{Imperfect conductivity}
\label{sec:conductivity}
{Throughout this  {paper}, we assumed that the objects are perfect reflectors {for the EM field}. In this section, we estimate  to what extent this is {justified} for realistic conductors.
Our analysis is based on the fact that the contribution to the Casimir force between two objects is suppressed at wave numbers greater than  their {inverse} separation.  Small frequencies contribute most, just where the assumption of perfect conductivity is best.  Although the distance scale for the wedge-plate and cone-plate geometries is $d$ --- the tip (or edge) to plate separation {---} the typical separations that contribute to the Casimir force for these {\it open} geometries is much greater than for parallel plates.

To investigate this {further,} we write the Casimir energy as an integral over (imaginary) wave number,
\begin{equation}\label{integral over energy density}
  \mathcal E =\int_{0}^{\infty} d\kappa \ \rho(\kappa,d)\,,
\end{equation}
and compare the Casimir energy density, $\rho(\kappa,d)$, for the cone-plate geometry with the parallel plate case.}
For a {\it perfectly} {reflecting} needle opposite an infinite plate, this density function can be read off the first line of Eq. (\ref{electromagnetic Needle}),
\begin{equation}
\rho(\kappa, d)\sim C \int_{1}^{\infty} dp \ \frac{p-1}{p+1} \frac{1}{p} e^{- 2 \kappa d p }\,,
\end{equation}
{where $C$ is a constant which depends on the opening angle.}
{Plots of the energy density are given in Fig.~\ref{rho}.} It is clear that the energy density of a needle-plate system is {falling off rapidly as opposed to} parallel plates where it falls off rather slowly.

\begin{figure}[ht]
\centering
\includegraphics[scale=.8]{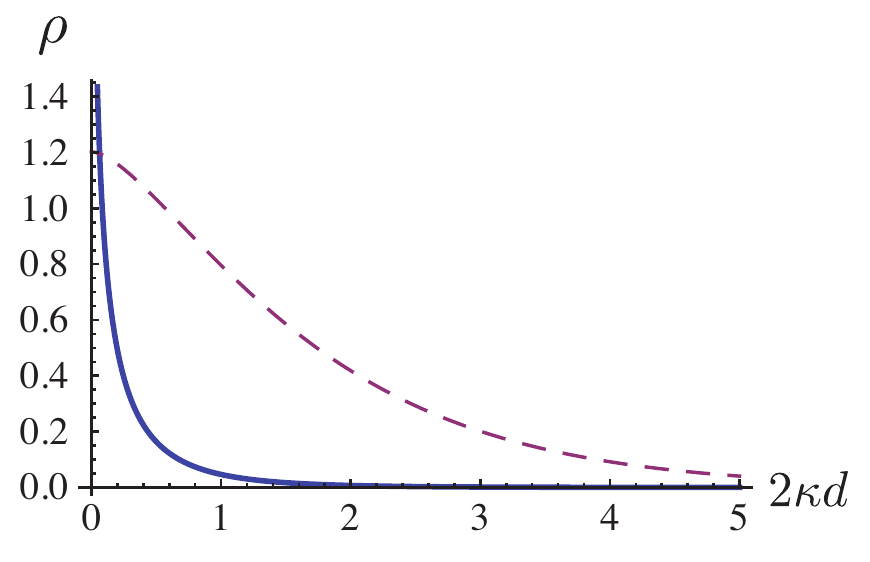}
\caption{Energy density $\rho(\kappa, d)${.
{The curves are not normalized and should be scaled with appropriate parameters of cone or plate geometry.}
The dashed line represents the energy density for two perfect metal plates, the solid line the energy density for a perfect metal cone and a plate. It is obvious that small frequencies are more important than large ones for the cone-plate as compared to the two-plate case.}}
\label{rho}
\end{figure}

{To be more quantitative, we  examine the Casimir energy as a function of the upper limit on the wave number integral,}
\begin{equation}
  \mathcal E(\kappa,d) =\int_{0}^{\kappa} d\kappa' \rho(\kappa',d)\,,
\label{Edepuplim}
\end{equation}
{and from this define a maximum wave number, $\kappa_{\rm max}$, above which the contribution to the Casimir force is negligible.}
The plots of $\mathcal E(\kappa,d)$ normalized by $\mathcal E(\infty,d)$ are given in Fig.~\ref{E} for the same two systems.
The cutoff, $\kappa_{\rm max}$, is defined by
\begin{equation}
\mathcal E(\kappa_{\rm max},d)=95\% \ \mathcal E(\infty,d),
\end{equation}
which gives an upper bound on the frequency at the price of 5\% inaccuracy.
The cutoff can be read off the plots which give $\kappa_{\rm max} \approx 0.5/ d $ for a needle-plate system and $\kappa_{\rm max} \approx 2/d$ for two parallel plates. The cutoff for the cone-plate system is considerably smaller, as anticipated.

\begin{figure}[ht]
\centering
\includegraphics[scale=.8]{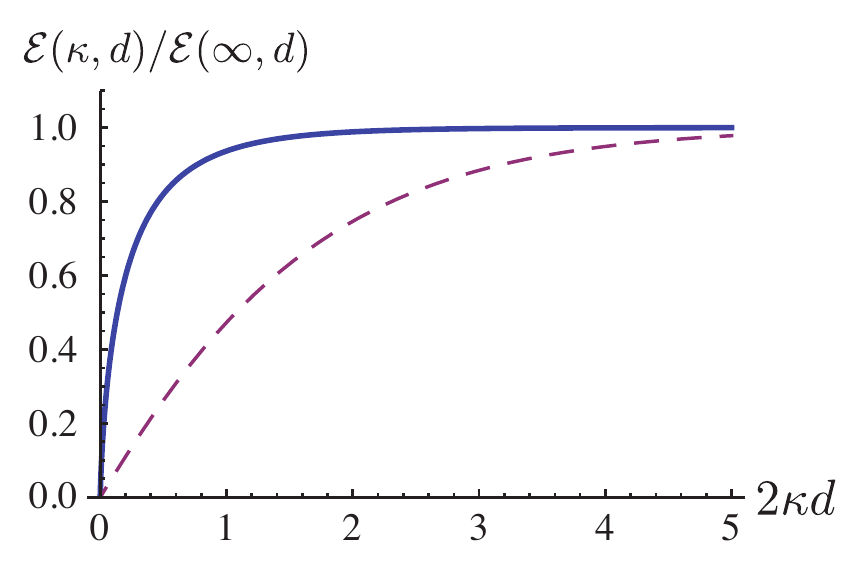}
\caption{Energy $\mathcal{E}(\kappa,d)${, normalized by $\mathcal{E}(\infty,d)$ (where, as usual, all frequencies are integrated over), is plotted} {as {a} function of $\kappa d$.} { The dashed line pertains to two perfect metal plates, the solid line to a perfect metal cone and plate. The energy in the cone-plate configuration is dominated by contributions from small frequencies when compared to the two-plate case.}}
\label{E}
\end{figure}

So far, we considered a perfect reflector and {required} that 95\% of the energy is contributed by frequencies smaller than $\kappa_{\rm max}$. Next we address the question  whether this regime can be approximated by perfect reflectivity.
A finite dielectric {function} only affects the $T$-matrix. Here, we derive an integral equation which determines the $T$-matrix and allows us to study the deviation from perfect reflectivity in a systematic way.

{In} the limit of a sharp cone, it suffices to consider only the $m=0$ components of the field. There is no $R$ mode in this sector. Hence, we need to solve the matrix equation for the $T$-matrix only for electromagnetic $M$ and $N$ modes at $m=0$. A further simplification occurs   for $m=0$ since the {polarizations} decouple. Finally to   lowest order in the opening angle, the $M$ mode can be neglected, as it can for the case of perfect reflectors. In summary, we must compute the scattering of the $N$ mode only at $m=0$.

By imposing the {continuity} conditions for electromagnetic field, we find
{for the $T$-matrix elements with $m=0$ the integral equation}
\begin{align}\label{Elec Tmat eqn}
  \int_{0}^{\infty} d \beta \ T_{\alpha \beta} \ & d_{\beta \gamma} \ \big(P_{i\beta-1/2}(-\cos\theta_0) \partial_{\theta_0} P_{i\gamma-1/2}(\cos \theta_0) -\frac{1}{\epsilon}\partial_{\theta_0} P_{i\beta-1/2}(-\cos\theta_0) P_{i\gamma-1/2}(\cos \theta_0)\big) \cr
    =&d_{\alpha \gamma} \ \big(P_{i\alpha-1/2}(\cos\theta_0) \partial_{\theta_0} P_{i\gamma-1/2}(\cos \theta_0)-\frac{1}{\epsilon}\partial_{\theta_0} P_{i\alpha-1/2}(\cos\theta_0) P_{i\gamma-1/2}(\cos \theta_0)\big) \,,
\end{align}
where $\epsilon$ is the permittivity and
\begin{equation} \label{d-eqn}
 d_{\beta \gamma}=  \int_{0}^{\infty} dx \ k_{i\beta-1/2}( x) k_{i\gamma-1/2}( n x),
\end{equation}
with $n=\sqrt\epsilon$. If $\epsilon$ is large, we can systematically organize the perturbation theory as
\begin{equation}
  T= T^{0}+T^{1}+ \cdots,
\end{equation}
where each order is smaller by a factor of $1/\epsilon$ and $T^0$ is the perfect reflector result.
Using this ansatz in the matrix equation and exploiting the smallness of $\theta_0$, we obtain
\begin{align}
      T^1_{\alpha \beta}=-  \frac{1}{\epsilon(i{c}\kappa)} \frac{2\pi}{\theta_0^2|\ln\theta_0/2|^2} \frac{1}{\cosh \beta\pi} \int d\gamma \ d_{\alpha \gamma }\frac{1}{\gamma^2+1/4} \ (d^{-1})_{\gamma \beta}.
\end{align}
Although we will not compute this integral explicitly, it can be compared parametrically with the zeroth order result, $T^0\sim 1/|\ln\theta_0/2|$. The first order solution is smaller than the perfect reflector result parametrically by a factor of
\begin{equation}
  \delta\equiv\frac{1}{\epsilon(i c\kappa)\ \theta_0^2|\ln\theta_0/2|}.
\end{equation}
Of course, for extremely small angles, perfect reflectivity is incorrect. However, if {the permittivity} $\epsilon$ is large enough, the correction can be small, even at small angles.
Note that, at worst, the frequency in the argument of $\epsilon$ should be replaced by $\kappa_{\rm max}$, which is given by $0.5/d$ for a sharp cone. Using the plasma model
\begin{equation}
  \epsilon(i {c} \kappa)=1+\frac{(2 \pi)^2}{(\lambda_p \kappa)^2} \,,
\end{equation}
for gold ($\lambda_p=137\rm nm$), and choosing a relatively small angle $\theta_0=0.1$ at a fairly small separation distance of $d=200\rm nm$, we get $\delta\approx 5\%$.
The correction to the {cone's} $T$-matrix implies the same correction to the energy ({in the first reflection,} the energy is just proportional to the $T$-matrix). Therefore, the assumption of perfect reflectivity seems to be quite good even for small angles and short distances.
{Here, we studied the finite conductivity corrections only for the cone to estimate its importance. The same effects for the infinite plate can be found exactly within this formalism since the $T$-matrix of a dielectric plate remains diagonal. By employing the same dielectric model at the same separation distance ($d=200nm$), we can easily see that finite conductivity correction for an infinite plate (facing a perfect conducting needle) is just a few percent.}\\

\section*{Appendix: The electromagnetic Green's function}
\label{AppendixGreenFn}

In this section we derive the electromagnetic Green's function in
coordinates appropriate to the cone problem.  Similar, though more
tedious, algebra will reproduce Eq. (\ref{ConvMatrixElecCone}),
which gives the conversion matrices.

We start from the dyadic Green's function in terms of spherical vector
wave functions,
\begin{align}
\mathbb{G}_0(\x;\x',i\kappa) & = \kappa
\sum_{l=1}^{\infty}\sum_{m=-\infty}^{\infty}
\curl i_l(\kappa r_<) Y_l^m(\mathbf{\hat{r}}) \x
\otimes
\curlp k_l(\kappa r_>) Y_l^{m*}(\mathbf{\hat{r}'}) \x'
\nonumber \\
& -
\frac{1}{\kappa}\curl\curl i_l(\kappa r_<) Y_l^{m}(\mathbf{\hat{r}}) \x
\otimes
\frac{1}{\kappa}\curlp\curlp k_l(\kappa r_>) Y_l^{m*}(\mathbf{\hat{r}'}) \x'\,,
\end{align}
and write this expression in a form which is more appropriate for our
purpose, as
\begin{align}\label{GnfnMMNNSph}
\mathbb{G}_0(\x;\x',i\kappa)= \sum_{l=0}^{\infty} \sum_{m=-l}^{l}
\frac{1}{l(l+1)} \frac{2 l+1 }{4 \pi} \left(\mathbf M_{l
m}^{\rm out}(\x')\otimes \mathbf{M}_{l m}^{{\rm reg}*}(\x)-\mathbf N_{l
m}^{\rm out}(\x')\otimes\mathbf{N}_{l m}^{{\rm reg}*}(\x)\right),
\end{align}
with
\begin{align}
 \mathbf  M_{lm}^{\rm reg}= \curl i_{l}(\kappa r ) P_{l}^{m}(\cos \theta)
 e^{i m \phi} \mathbf r, \nonumber \\
 \mathbf  N_{lm}^{\rm reg}= \frac{1}{\kappa}\curl \curl i_{l}(\kappa r )
 P_{l}^{m}(\cos \theta) e^{i m \phi} \mathbf r.
\end{align}
The outgoing wave functions are defined similarly, with
$k_\ell(\kappa r)$ substituted for $i_\ell(\kappa r)$. Then,
\begin{equation}
 \mathbf{ M}_{l m}^{\rm out}\otimes \mathbf{M}_{l m}^{{\rm reg}*}-\mathbf N_{l
 m}^{\rm out}\otimes \mathbf{N}_{l m}^{{\rm reg}*} = {\mathbb{D}} \left[ k_{l}(\kappa r')
 P_{l}^{m}(\cos \theta') e^{i m \phi'} (-1)^m i_{l}(\kappa r )
 P_{l}^{-m}(\cos \theta) e^{-i m \phi} \right],
\end{equation}
where the dyadic differential operator $\mathbb{D}$ is defined as
\begin{equation}
\mathbb{D} = \displaystyle \curl \vecr \otimes \curlp \vecr
'-\frac{1}{\kappa}\curl\curl \vecr \otimes \frac{1}{\kappa}\curlp \curlp
\vecr '.
\end{equation}
The summation over $m$ in Eq. (\ref{GnfnMMNNSph})
can be extended to $(-\infty,\infty)$ because
\begin{equation}
 P_{l}^{m}(\cos \theta') P_{l}{^{-m}}(\cos \theta)=0, \hskip .3in
\mbox{for $l < |m|$ and $l,m$ integers.}
\end{equation}
By using $P_{l}^{m}(\cos \theta)=(-1)^{l-m} P_{l}^{m}(-\cos \theta)$,
we can write the Green's function as
\begin{equation}
  \mathbb{G}_0=  \mathbb{D}\left[ \kappa \sum_{l=1}^{\infty}
  \sum_{m=-\infty}^{\infty} \frac{1}{l(l+1)} \frac{2 l+1 }{4 \pi}
  (-1)^l \ k_{l}(\kappa r') P_{l}^{m}(-\cos \theta') e^{i m \phi'} \
  i_{l}(\kappa r ) P_{l}^{-m}(\cos \theta) e^{-i m \phi} \right]\,.
\end{equation}
In order to turn the sum into a contour integral, we should divide it
by $\sin \nu \pi$ to obtain poles at the integers $\nu=l$. However,
unlike the scalar case, we now have to exclude $l=0$ from the sum.
Moreover, the factor of $1/l$ in front of the sum defines a pole at
the origin. Therefore, analytic continuation of the integral yields
two different contributions: an integral from $-1/2-i \infty$ to $-1/2+i
\infty$ denoted by $\mathcal C'$, and a small circle at the origin denoted by $\mathcal C_0$ (see Fig.~\ref{ConeElecCP}).  From the asymptotic
behavior of the Legendre functions given in Eq.
(\ref{LegendreAsymptotics}), we see that the large half-circle at
infinity can be neglected for $\theta'> \theta$.  So we have
\begin{equation} \label{GrnTwoPieces}
  \mathbb{G}_0=\mathbb{G}_0^{\mathcal{C}'}+\mathbb{G}_0^{\mathcal{C}_0},
\end{equation}
where
\begin{align}
  \mathbb{G}_0^{\mathcal{C}'}=
   - \frac{\kappa}{4 \pi}\sum_{m=-\infty}^{\infty}  \int_{0}^{\infty}
d\lambda  \ \lambda \tanh(\lambda
\pi)\frac{1}{\lambda^2+1/4}(\mathbf{M}_{i \lambda-1/2, m }^{\rm out}
\mathbf{M}_{i \lambda-1/2, m }^{{\rm reg}*} -\mathbf{N}_{i \lambda-1/2, m
}^{\rm out}\mathbf{N}_{i \lambda-1/2, m }^{{\rm reg}*})\,,
\end{align}
with the vector wavefunctions defined in Eq. (\ref{El Fns}), and
the residue of the pole at the origin gives
\begin{equation}
\mathbb{G}_0^{\mathcal{C}_0}=
\mathbb{D} \left[\oint_{\mathcal{C}_0} d \nu \frac{1}{\nu (\nu +1)}\frac{2 \nu+1}{4
\pi} \frac{\pi}{\sin \pi \nu} k_{\nu}(\kappa r') P_{\nu}^{m}(-\cos
\theta') \ i_{\nu}(\kappa r ) P_{\nu}^{-m}(\cos \theta) e^{-i m(\phi-
\phi')}\right] \, .
\end{equation}
{Because of the double pole in the integrand, we should take the first derivative of its coefficient (with respect to $\nu$). This is {further} simplified by noting that
$P_{0}^{m}(-\cos\theta') P_{0}^{-m}(\cos \theta)=0$ (except
when $m=0$, {which does not} contribute to the Green function since it vanishes upon applying the
${\mathbb{D}}$ operator). So we must compute }
\begin{equation}
  \partial_\nu \left( P_{\nu}^{m}(-\cos \theta') P_{\nu}^{-m}(\cos
  \theta)\right) \Big|_{\nu=0}.
\end{equation}
A bit of algebra gives
\begin{equation}
\mathbb{G}_0^{\mathcal{C}_0}  = \mathbb{D}\left[\frac{\kappa }{4\pi}
 \sum_{m=-\infty}^{\infty} k_{0}(\kappa r') i_{0}(\kappa r )
\Gamma(|m|)\Gamma(|m|+1) P_{0}^{-|m|}(-\cos \theta') \ P_{0}^{-|m|}(\cos
\theta) e^{-i m(\phi- \phi')} \right].
\end{equation}
This is not yet in the form that we need, because the radial
coordinates $r$ and $r'$ should appear symmetrically.  The symmetry
becomes manifest by applying {$\mathbb{D}$}. The double and
quadruple curls in the {$\mathbb{D}$} operator combine to yield
\begin{equation}
\mathbb{G}_0^{\mathcal{C}_0}=   - \frac{\kappa}{4 \pi}\sum_{m=-\infty,
m \ne 0}^{\infty} \Gamma(|m|)\Gamma(|m|+1) \mathbf{R}_{0, m}^{\rm out}\
\mathbf{R}_{0,m}^{{\rm reg}*}
\end{equation}
where {the} vector functions $R$ are defined in Eq.
(\ref{El Fns}).  We thus obtain the full Green's function, Eq.
(\ref{El Grn Fn}).\\

\section*{Acknowledgements}
   This work was supported by the National Science Foundation (NSF)
through grants PHY08-55426 (NG), DMR-08-03315 (SJR and MK),
Defense Advanced Research Projects Agency (DARPA) contract
No. S-000354 (SJR, MK, and TE), by the Deutsche Forschungsgemeinschaft
(DFG) through grant EM70/3 (TE), and by the U. S. Department of Energy
(DOE) under cooperative research agreement \#DF-FC02-94ER40818 (MFM, RLJ). We wish to thank Umar Mohideen for discussions on AFM tip.

\bibliographystyle{aip}

\end{document}